\documentclass[iop]{emulateapj}

\usepackage{color, colortbl}
\usepackage{rotating}
\usepackage{txfonts}
\usepackage{graphicx}
\usepackage{subfigure}
\usepackage{multirow}
\usepackage{verbatim}
\usepackage{longtable}
\usepackage{url}
\usepackage{float}
\usepackage{ esint }
\usepackage{dcolumn}
\usepackage{hyperref}
\usepackage[utf8]{inputenc}
\makeatletter
\newcommand\footnoteref[1]{\protected@xdef\@thefnmark{\ref{#1}}\@footnotemark}
\makeatother

\slugcomment{Accepted to AJ}

\shorttitle{\textit{Gaia} Binary Recoverability}
\shortauthors{Ziegler et al.}

\begin{document}

\title{Measuring the Recoverability of Close Binaries in $Gaia$ DR2 with the Robo-AO \textit{Kepler} Survey}

\author{Carl Ziegler\altaffilmark{1}, Nicholas M. Law\altaffilmark{2}, Christoph Baranec\altaffilmark{3}, Tim Morton\altaffilmark{4}, Reed Riddle\altaffilmark{5},  Nathan De Lee\altaffilmark{6,7}, Daniel Huber\altaffilmark{3,8,9,10}, Suvrath Mahadevan\altaffilmark{11,12}, Joshua Pepper\altaffilmark{13}}

\email{carl.ziegler@dunlap.utoronto.ca}
\altaffiltext{1}{Dunlap Institute for Astronomy and Astrophysics, University of Toronto, Ontario M5S 3H4, Canada}
\altaffiltext{2}{Department of Physics and Astronomy, University of North Carolina at Chapel Hill, Chapel Hill, NC 27599-3255, USA}
\altaffiltext{3}{Institute for Astronomy, University of Hawai`i at M\={a}noa, Hilo, HI 96720-2700, USA}
\altaffiltext{4}{Department of Astrophysical Sciences, Princeton University, Princeton, NJ 08544, USA}
\altaffiltext{5}{Division of Physics, Mathematics, and Astronomy, California Institute of Technology, Pasadena, CA 91125, USA}
\altaffiltext{6}{Department of Physics, Geology, and Engineering Technology, Northern Kentucky University, Highland Heights, KY 41099, USA}
\altaffiltext{7}{Department of Physics and Astronomy, Vanderbilt University, Nashville, TN 37235, USA}
\altaffiltext{8}{Sydney Institute for Astronomy (SIfA), School of Physics, University of Sydney, NSW 2006, Australia}
\altaffiltext{9}{SETI Institute, 189 Bernardo Avenue, Mountain View, CA 94043, USA}
\altaffiltext{10}{Stellar Astrophysics Centre, Department of Physics and Astronomy, Aarhus University, Ny Munkegade 120, DK-8000 Aarhus C, Denmark}
\altaffiltext{11}{Department of Astronomy and Astrophysics, The Pennsylvania State University, University Park, PA 16802, USA}
\altaffiltext{12}{Center for Exoplanets and Habitable Worlds, The Pennsylvania State University, University Park, PA 16802, USA}
\altaffiltext{13}{Department of Physics, Lehigh University, 16 Memorial Drive East, Bethlehem, PA 18015, USA}
\begin{abstract}

We use the Robo-AO survey of \textit{Kepler} planetary candidate host stars, the largest adaptive optics survey yet performed, to measure the recovery rate of close stellar binaries in \textit{Gaia} DR2. We find that \textit{Gaia} recovers binaries consistently down to 1\arcsec and at magnitude contrasts as large as 7; close systems are not often not resolved, regardless of secondary brightness. \textit{Gaia} DR2 binary detection does not have a strong dependence on the orientation of the stellar pairs. We find 177 nearby stars to \textit{Kepler} planetary candidate host stars in \textit{Gaia} DR2 that were not detected in the Robo-AO survey, almost all of which are faint ($G>$20); the remainder were largely targets observed by Robo-AO in poor conditions. If the primary star is the host, the impact on the radii estimates of planet candidates in these systems is likely minimal; many of these faint stars, however, could be faint eclipsing binaries that are the source of a false positive planetary transit signal. With Robo-AO and \textit{Gaia} combined, we find that 18.7$\pm$0.7\% of \textit{Kepler} planet candidate hosts have nearby stars within 4\arcsec. We also find 36 nearby stars in \textit{Gaia} DR2 around 35 planetary candidate host stars detected with K2. The nearby star fraction rate for K2 planetary candidates is significantly lower than that for the primary \textit{Kepler} mission. The binary recovery rate of \textit{Gaia} will improve initial radius estimates of future TESS planet candidates significantly, however ground-based high-resolution follow-up observations are still needed for precise characterization and confirmation. The sensitivity of \textit{Gaia} to closely separated binaries is expected to improve in later data releases.
\end{abstract}


\section{Introduction}

The \textit{Gaia} Data Release 2 (DR2) has provided astrometry, parallaxes, and photometry for over a billion stars in the galaxy \citep{gaiadr2}. Many of these stars are in fact close binaries: approximately half of solar-type stars form with at least one companion \citep{raghavan10, moe17}. Understanding the multiplicity of stellar populations can provide insight into various stellar formation processes and evolution scenarios \citep{zhang13, ziegler15}, as well as provide constraints for theoretical models and mass-luminosity relationships \citep{chabrier00}. The presence of a previously unknown stellar companion to a transiting-planet-hosting star can substantially increase the estimate of the radius of planets due to the additional flux from the non-transited star \citep{ciardi15, ziegler18a}. The Transiting Exoplanet Survey Satellite \citep[TESS,][]{ricker15}, with detector pixels $\sim$25$\times$ the size of \textit{Kepler}, will be particularly susceptible to contamination from nearby sources. In addition, there is significant evidence that stellar binaries can sculpt \citep{ziegler18b} or disrupt \citep{kraus16} planetary systems. Many bound systems have sub-arcsecond separations \citep{ziegler18b} and currently require high-angular resolution instruments on the ground to detect.

With a primary mirror 1.45-m in size in the scanning direction \citep{gaia2016b}, the ability of \textit{Gaia} to resolve close binaries should be comparable to the \textit{Hubble Space Telescope}. \textit{Gaia} Data Release 1 was limited to angular resolutions of 2-4\arcsec due to data processing limitations \citep{arenou17}. DR2 greatly improved on this, sensitive to most $>2$\arcsec pairs, but only a small fraction of sub-arcsecond pairs were resolved \citep{arenou18}. The probability that \textit{Gaia} will resolve stellar binaries is not solely a function of separation, however, but also of the flux ratio of the pair and, due to the rectangular pixels of \textit{Gaia} induced by the scanning direction, the position angle between the two stars \citep{debruijne15}. The close binaries not resolved in DR2 are handled as single objects, with blended photometry and occasional spurious astrometric solutions \citep{arenou18}.

There is also the potential for spurious source detections in \textit{Gaia} DR2. The dominant source of these detections is from diffraction spikes around stars brighter than 16 mag \citep{dr1_sourcelist}. Many of these spurious detections are identified by comparing data from multiple transits (i.e., checking whether the source is consistent in subsequent observations). A fraction of these erroneous detections (less than 20\%) remained in the \textit{Gaia} Data Release 1, with DR2 expected to be significantly cleaner \citep{gaiadr2}.

The Robo-AO \textit{Kepler} survey, the largest adaptive optics survey yet performed, with 3857 planetary candidate host stars observed, is an excellent test of the recovery rate of binaries in \textit{Gaia} DR2. Robo-AO, the first autonomous adaptive optics instrument, detected 620 companions\footnote{For brevity we denote stars which we found within our detection radius of KOIs as ``companions,'' in the sense that they are asterisms associated on the sky. For more on the probability of association between each pair of stars, see \citet{ziegler18}.} at separations between 0\farcs15 and 4\farcs0 and at contrasts up to 7 mags \citep{law14, baranec16, ziegler18a, ziegler16}. The set of \textit{Kepler} planet candidates host stars are largely 12$<$$G$$<$17, a brightness regime nearly complete in DR2 \citep{arenou18}, and detected companions down to the \textit{Gaia} faint limit ($G$$\approx$21). With this large homogeneous set of high-angular resolution observations, the ability of \textit{Gaia} to recover binaries as a function of separation, contrast, and orientation can be finely quantified.

We begin in Section \ref{sec:methodology} by briefly describing the Robo-AO system and the Robo-AO observations of \textit{Kepler} planetary candidates. We then describe the crossmatching of the Robo-AO detections with the \textit{Gaia} DR2 catalog. We present and discuss the results in Section \ref{sec:results}, including the implications for future transiting planet surveys, and conclude in Section \ref{sec:conclusion}.

\section{Methodology}
\label{sec:methodology}

\subsection{Robo-AO Observations}
Observations in the survey were performed using the Robo-AO automated laser adaptive optics system at Palomar and Kitt Peak \citep{baranec14, baranec17, jc18} that can efficiently perform large, high angular resolution surveys. The adaptive optics system runs at a loop rate of 1.2 kHz to correct high-order wavefront aberrations, delivering median Strehl ratios of 9\% and 4\% in the \textit{i}\textsuperscript{$\prime$}-band at Palomar and Kitt Peak, respectively. Observations were between 90 and 120 s, and taken in a long-pass filter cutting on at 600 nm. The LP600 filter approximates the \textit{Kepler} passband at redder wavelengths, while also suppressing blue wavelengths that reduce adaptive optics performance. The LP600 passband is compared to the \textit{Kepler} passband in Figure 1 of \citet{law14}. We obtained high-angular-resolution images of 3313 KOIs with Robo-AO between 2012 July 16 and 2015 June 12 (UT) at the Palomar 1.5m telescope. We observed 532 additional KOIs with Robo-AO between 2016 June 8 and 2016 July 15 (UT) at the Kitt Peak 2.1m telescope.

\subsection{Gaia-Kepler Crossmatching}

The positions of the \textit{Kepler} planetary candidates \citep{dr25, thompson18} were cross-matched on the \textit{Gaia} online archive service\footnote{http://gea.esac.esa.int/archive/} with an advanced Astronomical Data Query Language (ADQL) search. This provided a list of sources in \textit{Gaia} DR2 within 5\arcsec of each planet candidate host star. To identify the likely primary star in multiple systems, we applied a magnitude cut using the \textit{Kepler} magnitude of the host star and the \textit{Gaia G}-magnitude of each source. The star with a \textit{G}-magnitude within 1 magnitude of the host star's \textit{Kepler} magnitude was determined to be the primary star. If multiple stars had nearly equivalent brightness, or if no star had a magnitude similar to that in the \textit{Kepler} catalog, the closest star to the coordinates of the planet candidate host star was determined to be the primary star. In general, the coordinates of the primary star were within 0\farcs20 of the positions reported in the \textit{Kepler} catalog. Several planet candidate host stars had no clear source in \textit{Gaia} DR2: KOI-98, 227, 640, 959, 1152, and 6728. These systems have been excluded from this analysis.

We searched for potential spurious detections in our crossmatch using the \textit{Gaia} parallaxes and distance solutions of \citet{bailerjones18}. We found no sources with distances less than 1 pc possibly originating in the solar system or greater than 20 kpc extra-galactic in our sample. Likewise, none had parallaxes greater than 1\arcsec or less than 0.05 mas. None of our sources were found in the catalog of known solar system objects \citep{sso_gaia}. Lastly, the majority of the stars have magnitudes in two additional photometric bands (BP in the blue, and RP in the red) obtained from integrating the \textit{Gaia} prism spectra. All of the planet candidate hosts and nearby stars with the available photometry had reasonable colors (-1$<$(BP-RP)$<$4), consistent with that of a stellar source \citep{andrae18}.

The separation and contrast of any additional sources detected in the area of sky around each host star were compared to the companion properties measured by Robo-AO. The Robo-AO observations were performed between 2012 and 2016, and the positions of the primary and secondary stars have likely shifted with respect to the \textit{Gaia} reference epoch (J2015.5). We used the positions and proper motion of the stars detected by \textit{Gaia} to determine their positions when the Robo-AO observations were performed, using the Astropy software package \citep{astropy}.

The complete list of detections of nearby stars to planet candidate host stars is available in Table$~\ref{tab:gaiadetections}$. Nearby stars in \textit{Gaia} DR2 with similar contrasts and separations (\textit{G}-magnitude within 1 magnitude and separations within 0\farcs20) to the nearby stars detected using Robo-AO were classified as ``recovered," and nearby stars detected with Robo-AO that are not in \textit{Gaia} DR2 were classified as ``not-recovered." We also search for nearby stars in the \textit{Gaia} DR2 catalog that were not detected by Robo-AO, and list these detections in Table$~\ref{tab:newgaiadetections}$ (systems with more than two stars have additional rows for each nearby star). The separation and position angle of these binaries were calculated using the \textit{Gaia} coordinates using the Astropy software package \citep{astropy}, and the magnitude contrast is calculated from the reported \textit{Gaia} magnitudes.

In addition to searching around planet candidates from the primary \textit{Kepler} mission, we also searched \textit{Gaia} DR2 for sources nearby planetary candidates identified from the ongoing K2 mission. We acquired a list of these planet candidates, 773 in total, and their positions, sourced from EPIC \citep{huber16}, from the NASA Exoplanet Archive\footnote{https://exoplanetarchive.ipac.caltech.edu/}. A list of sources from DR2 within 5\arcsec of the positions of the candidates was generated, and the host star was identified with magnitude cuts using the \textit{Kepler} magnitude and \textit{Gaia G}-magnitude. Detections of sources nearby K2 planet candidate host stars are listed in Table$~\ref{tab:newgaiadetectionsk2}$.

\section{Results and Discussion}
\label{sec:results}

\subsection{Properties of Recovered Stars}
We find that, of the 620 stars detected with Robo-AO within 4\arcsec of 3857 \textit{Kepler} planetary candidates, 484, or 78$\%$, appear in \textit{Gaia} DR2. The recovery classifications for each star is listed in Table$~\ref{tab:gaiadetections}$, along with the Robo-AO measured binary properties and \textit{Gaia} DR2 source IDs for the primary and recovered secondary stars.

\begin{figure}
\centering
\includegraphics[width=240pt]{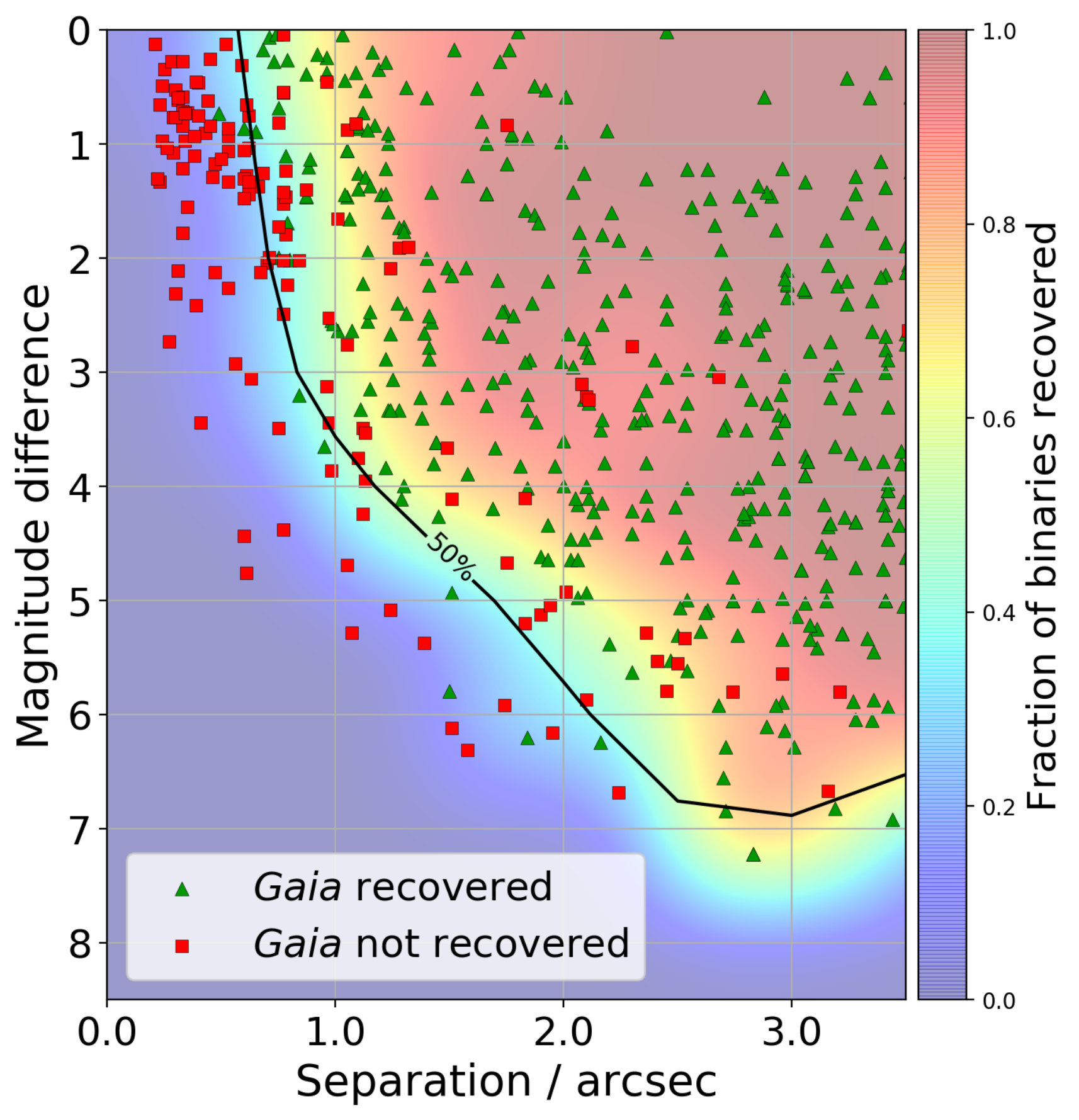}
\caption{The Robo-AO detections of nearby stars to \textit{Kepler} planetary candidates recovered and not-recovered in \textit{Gaia} DR2. The fraction of binaries recovered is plotted as a function of magnitude difference and separation from the primary star, calculated by measuring the number of recovered and non-recovered stars in bins of size 1-magnitude and 0\farcs5 and employing a bicubic interpolation. The 50\% recoverability contour has been labeled. In general, stars within 0\farcs75 of the primary star are not recovered in \textit{Gaia} DR2.}
\label{fig:heatmap_recovery}
\end{figure}

In general, most stars within 1\arcsec of the planetary candidate host star were not recovered (22.4\% recovery rate), and stars at separations greater than 2\arcsec were nearly all recovered (93\% recovery rate) down to the \textit{Gaia} faint limit. These recovery rates could potentially be influenced by the ability of Robo-AO to detect binaries at given separations and contrasts in some observations due to low-image performance, resulting from bad seeing or a faint target star. For magnitude contrasts less than 3, a region of high completeness for Robo-AO (companions at separations from 0\farcs15 to 4\arcsec are detectable in nearly all images), the recovery rate is 22.9\% within 1\arcsec, and 97\% at separations greater than 2\arcsec. For \textit{Kepler} planet candidate hosts, the majority of stars within 1\arcsec are members of likely bound stellar pairs, and their influence can have a significant impact on the architecture of the planetary system \citep{ziegler18b}. In Figure$~\ref{fig:heatmap_recovery}$, we plot the Robo-AO detections recovered and not-recovered by \textit{Gaia}, as well as the fraction of binaries recovered as a function of magnitude difference and separation.

\begin{figure}
\centering
\includegraphics[width=240pt]{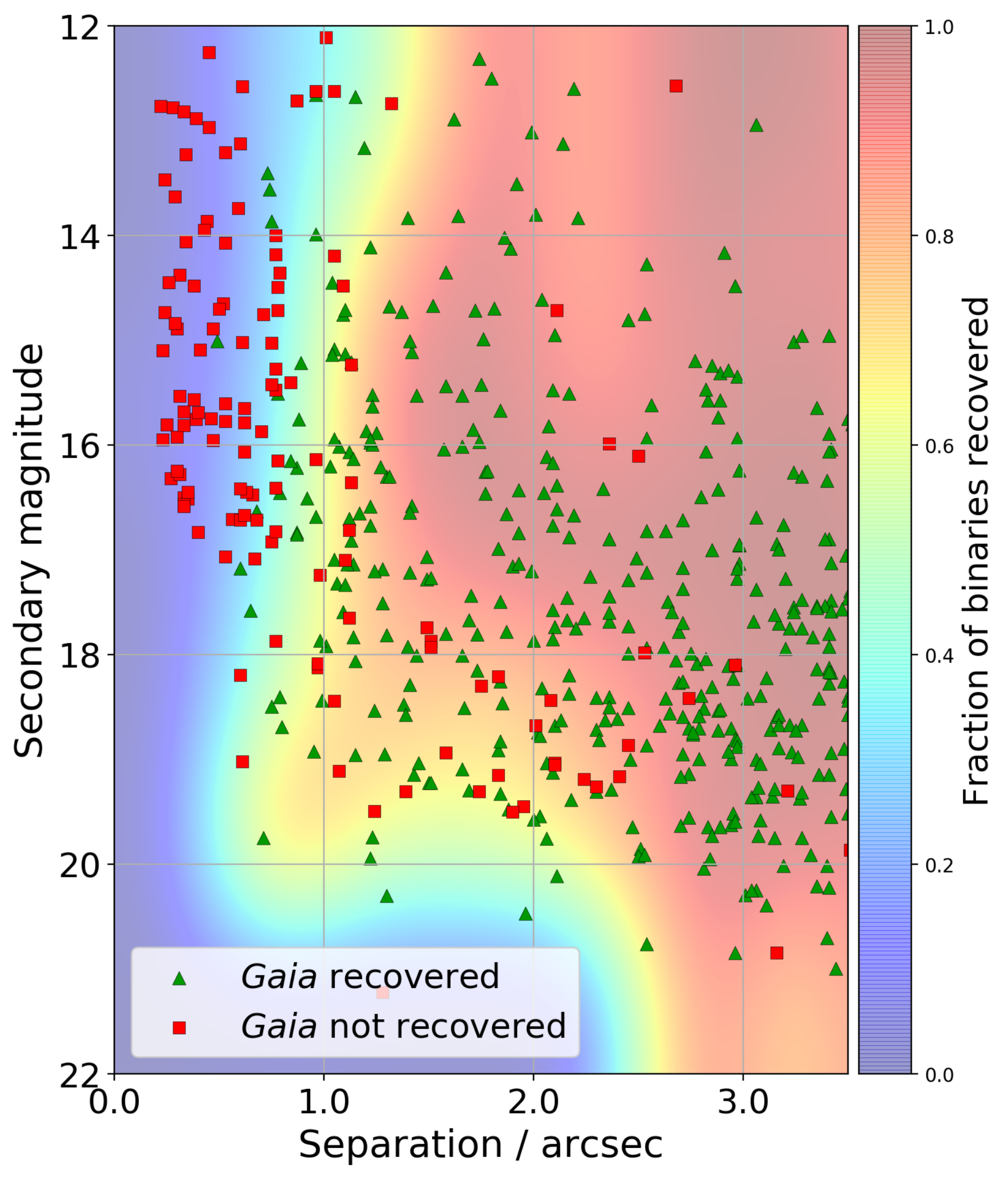}
\caption{Same as Figure $~\ref{fig:heatmap_recovery}$, however plotted as a function of the secondary star \textit{G}-magnitude. For recovered binaries, the secondary magnitude was measured by \textit{Gaia}; for non-recovered binaries, this magnitude is approximate and estimated using the primary star's \textit{G}-magnitude plus the visible contrast measured by Robo-AO.  Most stars within 1\arcsec are not recovered, and the recovery rate at low-separations is not dependent on secondary magnitude.}
\label{fig:heatmap_recovery_secondarymag}
\end{figure}

We also find that the recovery rate at low-separations does not depend on the brightness of the secondary star. In Figure$~\ref{fig:heatmap_recovery_secondarymag}$, we plot the fraction of binaries recovered as a function of the secondary star's magnitude and separation. We find that even at the bright end (m$_{G}$$<$13), very few stars are detected within 1\arcsec of the primary star.

\begin{figure}
\centering
\includegraphics[width=240pt]{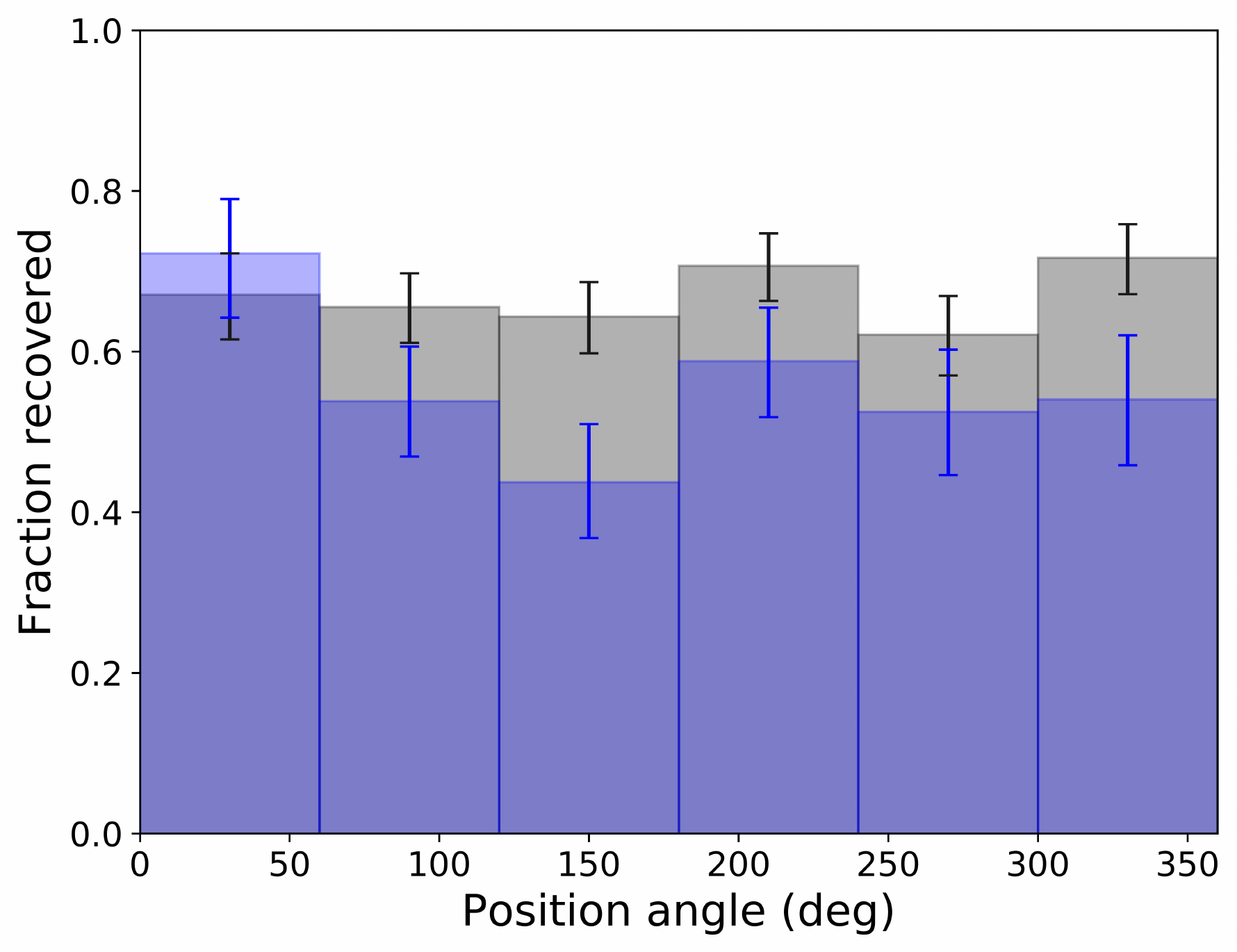}
\caption{The fraction of nearby stars recovered as a function of position angle with respect to the primary star for all Robo-AO detected stars within 4\arcsec and 2\arcsec in grey and blue, respectively. The recovery rate of nearby stars in \textit{Gaia} DR2 is not strongly dependent on the position angle of the stars.}
\label{fig:pa}
\end{figure}

The rectangular \textit{Gaia} pixels (with a 3:1 size ratio between across-scan and along-scan pixels) may introduce an orientation dependence to the ability of \textit{Gaia} to resolve close binaries \citep{debruijne15}. This asymmetric sensitivity is not expected to impact the final \textit{Gaia} catalog, as each object will be observed approximately 70 times at various orientations. However, it may be apparent in the recovery rate of binaries in the DR2 catalog, which is based on 22 months of data collection. In Figure$~\ref{fig:pa}$, we plot the fraction of stars detected with Robo-AO recovered in \textit{Gaia} DR2 as a function of position angle. The recovery rates in six position angle bins are all consistent with the overall recovery fraction. If we limit the set to only small-separation binaries ($\rho$$<$2\arcsec), as most of the variation in recovery will likely occur at these smaller separations, the recovery rate is consistent across all position angles.

\subsection{New Gaia Detections around Kepler Planet Candidates}

Within 4\arcsec of the 3857 \textit{Kepler} planetary candidate host stars observed by Robo-AO, \textit{Gaia} DR2 catalogs 177 nearby stars around 163 host stars that were not detected in the Robo-AO survey. The properties of these nearby stars, calculated from the \textit{Gaia} astrometry and photometry, are listed in Table$~\ref{tab:newgaiadetections}$. The majority of these detections fall outside of the sensitivity of Robo-AO, including nearly two-thirds (65\%) fainter than 20 mag. Longer integration times with Robo-AO could potentially observe some of these faint stars. We searched the Robo-AO images for any detection of a companion at the purported position of the nearby stars detected by \textit{Gaia} (accounting for proper motion shift). None were detected with 5$\sigma$ significance, however several low-significance detections were apparent to visual inspection. A future study using high-resolution data from a large-aperture telescope (such as Keck-AO) could potentially determine the validity of these faint \textit{Gaia} detections nearby bright stars.

\begin{figure}
\centering
\includegraphics[width=240pt]{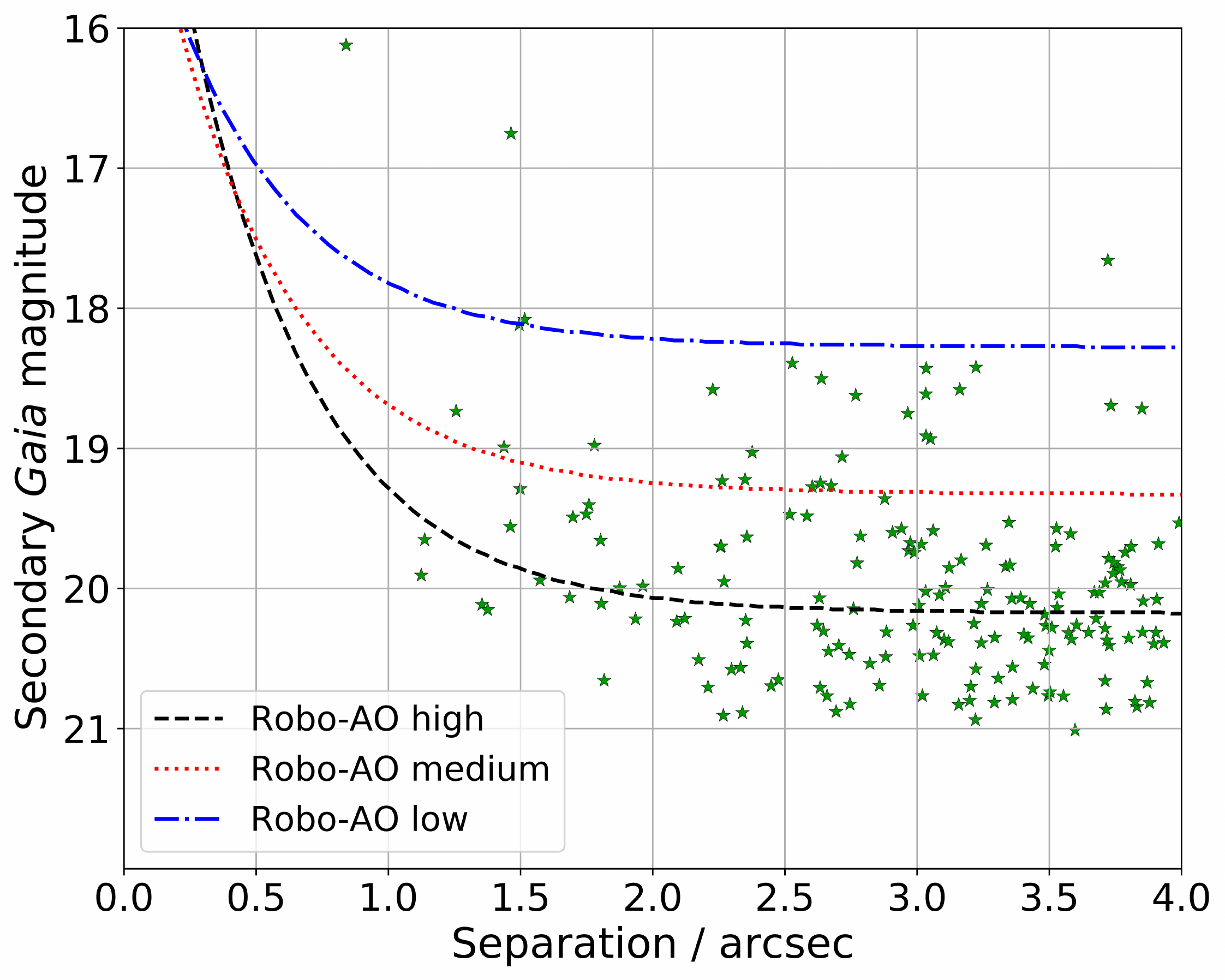}
\caption{Nearby stars to \textit{Kepler} planetary candidates in \textit{Gaia} DR2 that were not detected in Robo-AO images. Typical contrast curves for Robo-AO, in approximate \textit{Gaia g-}band magnitudes, are included for three image performance groups. The majority of these nearby stars were too faint for a significant detection in the Robo-AO images.}
\label{fig:newgaia}
\end{figure}

Altogether, approximately 99.5\% of secondary stars with $G<$18 detected by \textit{Gaia} were also detected in the Robo-AO \textit{Kepler} survey. The exceptions, all from particularly low-performance observations with shallow contrast curves, are secondaries in DR2 nearby KOI-118, 433, and 5736. The properties of the new detections are plotted in Figure$~\ref{fig:newgaia}$, along with typical Robo-AO visible-light contrast curves for three image performance groups, determined using the PSF core size as described in \citet{law14}.

The Robo-AO \textit{Kepler} survey found a nearby star fraction rate of 14.5$\pm$0.6\% in the Robo-AO detectability range (separations between $\sim$0$\farcs$15 and 4$\farcs$0 and $\Delta$m$\le$6). With the additional nearby stars in \textit{Gaia} DR2 combined with the Robo-AO detections, the nearby star rate of \textit{Kepler} planet candidate hosts is 18.7$\pm$0.7\%. Outside of 1\arcsec, where \textit{Gaia} recovers the majority of binaries, the nearby star fraction rate for Robo-AO and \textit{Gaia} is 11.3$\pm$0.5\%.

\subsection{Kepler planetary candidate radii}

A nearby star in the same photometric aperture as the target star will dilute the observed transit depth, resulting in underestimated radius estimates. In systems with a detected nearby star by Robo-AO, the estimated planetary radius will increase by a factor of 2.18, on average, if either star is assumed to be equally likely to host the planet \citet{ziegler18a}. For just systems with likely bound stars, determined with photometric parallaxes, the radii will increase by a factor of 1.77, on average \citep{ziegler18b}.

The nearby stars in \textit{Gaia} DR2 that were not detected by Robo-AO are, in general, faint and widely separated from the host star. Galactic simulations suggest that the majority of these stars are likely not bound to the primary star \citep{horch14,ziegler18b}. Assuming the planet indeed orbits the primary star, we use the relation from \citet{law14} to correct for the transit dilution,
\begin{equation}
R_{p,A}=R_{p,0}\sqrt{\frac{1}{F_{A}}}
\end{equation}
where R$_{p,A}$ is the corrected radius of the planet orbiting the primary star, R$_{p,0}$ is the original planetary radius estimate based on the diluted transit signal, and F$_{A}$ is the fraction of flux within the aperture from the primary star.

With the high contrasts of the newly detected \textit{Gaia} stars, their contamination of the \textit{Kepler} light curves is minimal. Using the \textit{Gaia} photometry as a proxy for the \textit{Kepler} photometry, if the transiting planet candidates orbit the primary star, their radii will increase by a factor of 1.007 due to the additional flux from these faint stars.

If instead, these planets orbit the secondary star, the corrected planet radius estimate relies on the radius of the secondary star, which is generally not known without color information. If we assume that all nearby stars are bound to the primary star, and use as the secondary radius the radius of an appropriately fainter star within the Dartmouth stellar models \citep{dotter08}, we can use the relation 
\begin{equation}
R_{p,B}=R_{p,0}\frac{R_{B}}{R_{A}}\sqrt{\frac{1}{F_{B}}}
\end{equation}
where R$_{p,B}$ is the corrected radius of the planet orbiting the secondary star bound to the primary star, R$_{B}$ and R$_{A}$ are the stellar radii of the secondary and primary star, respectively, and F$_{B}$ is the fraction of flux within the aperture from the secondary star. In this scenario, the planetary radii will increase by, on average, a factor of 8.2 in these systems. This scenario is unrealistic, however, and leads to a planetary population with a large fraction of gas giants, which is inconsistent with the understood planet occurrence rates of the galaxy \citep{howard12}. This scenario should be investigated for rare, difficult-to-model systems, such as those with unlikely dynamical properties, where one or more planet candidates could, in fact, be associated with the secondary star.

\subsection{Nearby Stars in Gaia to K2 Planet Candidates}

\begin{figure}
\centering
\includegraphics[width=240pt]{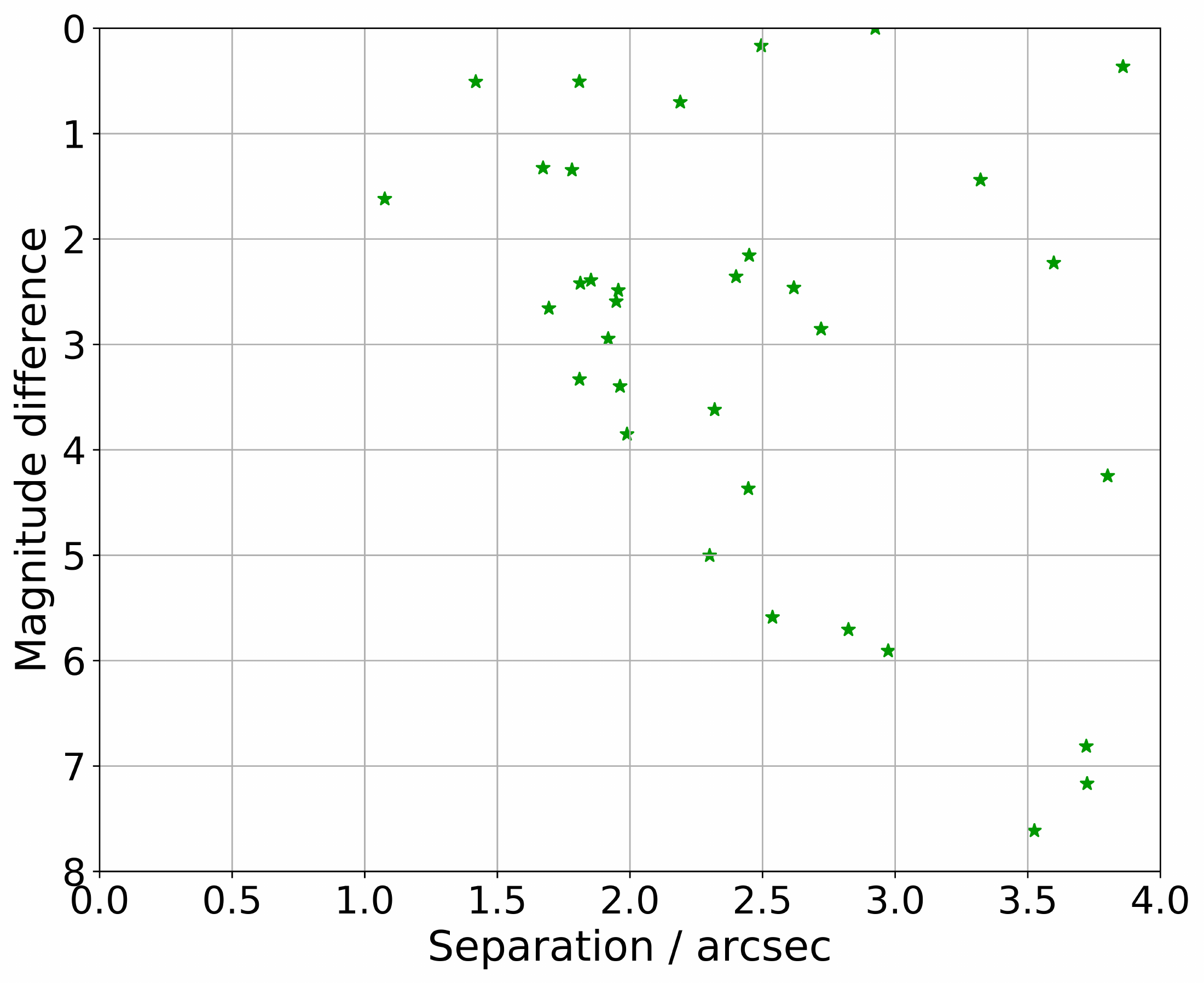}
\caption{Nearby stars to K2 planetary candidates in \textit{Gaia} DR2. The nearby star rate of K2 planet candidates is less than half that of \textit{Kepler} planet candidates in \textit{Gaia} DR2.}
\label{fig:k2}
\end{figure}

We searched for nearby stars in the \textit{Gaia} DR2 catalog around 773 K2 planet candidates from the first eight K2 campaigns, as listed on the NASA Exoplanet Archive. We found 36 nearby stars around 35 planet candidate hosts. The properties of these detected stars are listed in Table$~\ref{tab:newgaiadetectionsk2}$ and plotted in Figure$~\ref{fig:k2}$.

The fraction of nearby stars in \textit{Gaia} DR2 to K2 planetary candidates (4.5\%) is significantly lower than that of \textit{Kepler} planet candidates (9.7\%). The disparity between the nearby star fraction of \textit{Kepler} and K2 planet candidate hosts may be due to the K2 fields, which follow the ecliptic, appearing in less dense stellar regions with fewer unassociated background or foreground stars. The K2 targets lie, on average, at approximately $|b|$=38$^{\circ}$, far from the galactic plane and a significantly less dense region of the sky than the primary \textit{Kepler} mission (which had a center of field at $b$=14$^{\circ}$). In addition, \textit{Gaia} operates with a scanning law that passes through the north and south ecliptic poles every six hours, resulting in over twice as many observations at high ecliptic latitudes, such as the original \textit{Kepler} field, as at the ecliptic plane, where the K2 fields lie \citep{gaia2016b}. The additional observations likely improved the sensitivity of \textit{Gaia} to closely separated stars in the K2 fields \citep{debruijne15}.

Lastly, part of the disparity between the two samples may also in part be due to the larger fraction of late-type stars in K2 \citep{huber16}, which have, at these distances, an adaptive optics resolvable binarity rate of approximately half that of solar-type stars \citep{law05, janson12}. Indeed, only 2 of the 36 (5.5\%) nearby stars to K2 candidates in \textit{Gaia} DR2 lie at separations less than 1\farcs5, compared to 78 of 420 (17.3\%) for the \textit{Kepler} candidates, consistent with a low inherent binarity rate in the K2 sample.

\begin{table*}
\centering
\caption{\label{tab:newgaiadetectionsk2}Nearby Stars to K2 Planetary Candidates in \textit{Gaia} DR2}
\begin{tabular}{ccccccc}
\hline
\hline
\noalign{\vskip 3pt}  
Planet & Sep. & P.A. & $\Delta$m$_{G}$ & K2  & Primary \textit{Gaia} & Secondary \textit{Gaia} \\
candidate & (\arcsec) & (deg.) & (mags) & campaign & DR2 source ID & DR2 source ID \\
\hline \\ [-1.5ex]

202066537.01 & 2.19 & 74 & 0.7 & 0 & 3364627558065966848 & 3364627562364388352 \\ 
202086968.01 & 1.95 & 189 & 2.59 & 0 & 3369361402301215616 & 3369361406595494016 \\ 
201441872.01 & 3.86 & 240 & 0.37 & 1 & 3797258174978678016 & 3797258174978677888 \\ 
201546283.01 & 2.97 & 177 & 5.91 & 1 & 3798552815560689792 & 3798552811267494016 \\ 
201637175.01 & 1.92 & 226 & 2.95 & 1 & 3811002791880297600 & 3811002787586327040 \\ 
201650711.01 & 1.78 & 332 & 1.35 & 1 & 3812335125095532672 & 3812335125094701056 \\ 
201683540.01 & 1.99 & 203 & 3.85 & 1 & 3811900543124260480 & 3811900543123607552 \\ 
201828749.01 & 2.45 & 57 & 2.16 & 1 & 3909309851641800704 & 3909309851641320832 \\ 
203099398.01 & 1.96 & 64 & 2.49 & 2 & 6042368383828169728 & 6042368388127562752 \\ 
205029914.01 & 3.32 & 5 & 1.44 & 2 & 4131047326528868352 & 4131047326531704960 \\ 
205029914.01 & 3.72 & 178 & 7.17 & 2 & 4131047326528868352 & 4131047330825537792 \\ 
205040048.01 & 3.8 & 330 & 4.25 & 2 & 6245720108744660480 & 6245720104449034880 \\ 
205071984.03 & 3.72 & 346 & 6.81 & 2 & 4130539180358512768 & 4130539184653092352 \\ 
210625740.01 & 3.6 & 348 & 2.23 & 4 & 46432827015149184 & 46432827013380608 \\ 
210666756.01 & 2.4 & 212 & 2.36 & 4 & 49835540624946304 & 49835540624946560 \\ 
210958990.01 & 1.81 & 239 & 2.42 & 4 & 52752231438638080 & 52752235733602304 \\ 
211509553.01 & 1.96 & 328 & 3.4 & 5 & 605593554127479936 & 605593554127091200 \\ 
211694226.01 & 1.81 & 223 & 0.51 & 5 & 609915592602320896 & 609915596898129664 \\ 
211791178.01 & 1.67 & 347 & 1.33 & 5 & 659785149366912768 & 659785145072281600 \\ 
211978865.01 & 1.08 & 26 & 1.62 & 5 & 675557368789973632 & 675557364493662720 \\ 
212398508.01 & 2.32 & 255 & 3.62 & 6 & 3606357633269598464 & 3606357598909037696 \\ 
212428509.01 & 1.81 & 73 & 3.33 & 6 & 3606604782867769216 & 3606604782867769344 \\ 
212577658.01 & 1.42 & 12 & 0.51 & 6 & 3613738139429952768 & 3613738139430802816 \\ 
212628098.01 & 1.85 & 20 & 2.39 & 6 & 3624010185078481664 & 3624010189374241280 \\ 
212646483.01 & 1.69 & 213 & 2.66 & 6 & 3615089503644951168 & 3615089507940163072 \\ 
212661144.01 & 2.72 & 294 & 2.85 & 6 & 3615758251528391680 & 3615758251528391808 \\ 
212679181.01 & 1.24 & 210 & -0.41 & 6 & 3630190784752117504 & 3630190784751508480 \\ 
213920015.01 & 1.09 & 198 & -0.09 & 7 & 6764880018721513856 & 6764880018726114688 \\ 
214254518.01 & 3.53 & 29 & 7.61 & 7 & 6763711645882517504 & 6763711650189991552 \\ 
216114172.01 & 2.62 & 52 & 2.46 & 7 & 6768794138383822848 & 6768794142677222272 \\ 
216405287.01 & 2.54 & 254 & 5.59 & 7 & 4078733014274957184 & 4078733014252539008 \\ 
217149884.01 & 2.82 & 86 & 5.71 & 7 & 4082665245798510848 & 4082665245790571264 \\ 
217855533.01 & 2.49 & 92 & 0.17 & 7 & 4083005128025277952 & 4083005132333049472 \\ 
219256848.01 & 2.93 & 253 & 0.0 & 7 & 4088264543134042880 & 4088264543142623104 \\ 
219420915.01 & 2.45 & 266 & 4.37 & 7 & 4087971969962403840 & 4087971969954763264 \\ 
220192485.01 & 2.3 & 72 & 5.0 & 8 & 2534555412207560960 & 2534555416500291328 \\

\hline
\end{tabular}
\end{table*}

\citet{crossfield16} observed in high-resolution 164 of the candidate planets from K2 campaigns 0-4 using Keck-AO, Palomar PHARO/PALM-3000, LBT-LMIRCam, Gemini-NIRI, and Robo-AO. Within the separation range in which \textit{Gaia} has high binary recovery rate (1-4\arcsec), 22 nearby stars were detected around 20 planet candidate hosts, for a nearby star fraction rate of 12.2\%. Only 7 of the 20 multiple systems were detected by \textit{Gaia}: EPIC 201546283, 201828749, 202066537, 205029914, 210666756, 210958990, 203099398. This recovery rate (35\%) is significantly less than that for \textit{Kepler} planet candidates within the same separation range (82\%).

The reason for the low binary recovery rate of \textit{Gaia} DR2 compared to the high-resolution imaging in the K2 fields is unclear. \citet{arenou18} found that DR2 recovered significantly more close binaries in low-density fields, similar to the first five K2 campaign fields. The majority of the observations performed in \citet{crossfield16} were done in the NIR, with 10 of 13 of the binary systems not detected in \textit{Gaia} DR2 having contrasts greater than 5 magnitude. It is possible that the secondary stars in these systems are below the \textit{Gaia} faint limit in the visible. 

Unlike with the \textit{Kepler} planet candidates, the dilution from nearby stars detected with high-resolution imaging has already been taken into account in many of the K2 planet candidate's reported radii estimates (e.g., \citet{crossfield16}). In addition, the literature has significant variations in the planetary radius estimates of many K2 planets, particularly those around late-type stars. This is largely due to highly uncertain stellar parameters derived from photometry alone. Consequently, we do not report radius corrections for the K2 candidates with detected nearby stars in \textit{Gaia} DR2.

\subsection{Implications for TESS}

\begin{figure}
\centering
\includegraphics[width=220pt]{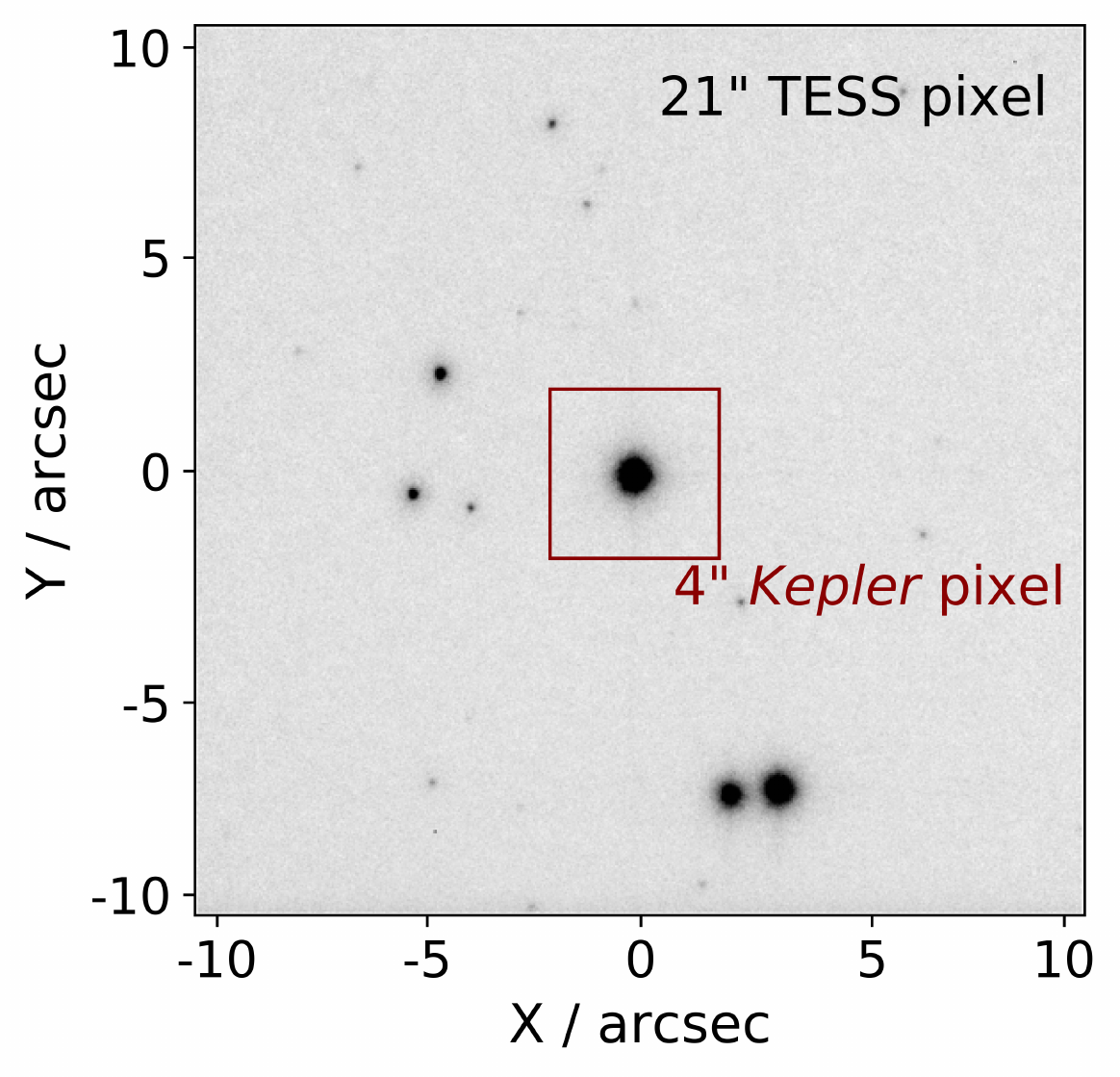}
\caption{A 21\arcsec square region of sky, the area subtended by a single TESS pixel, from a Robo-AO image centered on a super-Earth-sized planet candidate host, KOI-4725, located at \textit{b}=7.6. For comparison, the pixel size of \textit{Kepler} has been drawn. The transit signal from this planet candidate, if detected by TESS, would be diluted by multiple additional sources within the same photometric aperture (most TESS fields will, however, be in less crowded fields than the original \textit{Kepler} field). If not accounted for, the planetary candidate radius would be significantly underestimated due to this contamination; in this illustration, the planet candidate would exhibit a transit depth in uncorrected TESS data similar to an Earth-sized rocky planet. Each of these additional sources in this field is identified in \textit{Gaia} DR2.}
\label{fig:tesspixel}
\end{figure}

TESS, launched in April 2018, will search nearly the entire sky for transiting planets around bright, nearby stars \citep{ricker15}. Simulations estimate that TESS will detect over 10,000 exoplanets, including approximately 250 potentially rocky planets \citep{barclay18}. With significantly larger pixels than \textit{Kepler} (21\arcsec compared to 4\arcsec), the TESS light curves for most targeted stars will have some contamination from nearby stars (see Figure$~\ref{fig:tesspixel}$). In the case of a transiting planet, this additional flux dilutes the transit signal, resulting in underestimated planetary radii.

Ground-based, wide-field surveys, such as 2MASS or SDSS, typically detect near-equal contrast companions to within separations of 3\arcsec$~$ \citep{ziegler18a}. The recovery of nearby stars to \textit{Kepler} planet candidates proves that \textit{Gaia} DR2 is a far more complete census of the stellar population in the vicinity of TESS targets. \textit{Gaia} specifically is not sensitive to low-contrast, sub-arcsecond companions (although unresolved low-mass binaries may be identified, if not characterized, by their presence above the main-sequence using the precise parallaxes and stellar properties resulting from \textit{Gaia} DR2 \citep{berger18}). \citet{ziegler18a} found that for systems with Robo-AO detected nearby stars, the estimated radii of \textit{Kepler} planet candidates will increase by a factor of 1.54, on average, assuming the planet is equally likely to orbit the primary or secondary star. Using instead only the nearby stars detected by \textit{Gaia}, including those not detected by Robo-AO, the planet candidates radii estimates will increase by a factor of 2.47, under the same assumptions. Of course, the stars detected by \textit{Gaia} DR2 are, in general, much fainter and widely separated and are unlikely to be bound to the primary star \citep{horch14, ziegler18b}. The scenario in which the primary and secondary star are equally likely to host the star is not likely and leads to a high occurrence rate of Jupiter-sized planets that has not been observed \citep{howard12}. If instead, all planets orbit the primary star, the additional flux from the \textit{Gaia} detected stars will lead to the radii of planet candidates in multiple systems increasing by a factor of 1.12, on average.

With \textit{Gaia} DR2, the properties of a large number of nearby stars not resolved in seeing-limited ground-based surveys will be readily available, greatly improving the initial radius estimates of detected TESS planets. Ultimately, however, the TESS planet candidates will each require ground-based high-resolution follow-up observations to identify the close, likely bound stars, as well as provide more precise characterization and confirmation. With \textit{Gaia} DR2 alone, the radius estimates of 254 \textit{Kepler} planet candidates would be underestimated due to non-recovery of close binaries which could be detected with high-resolution instruments. Fortunately, the brightness of the TESS targets, typically 2-5 mag brighter than \textit{Kepler} targets, will allow smaller telescopes with less-costly high-angular resolution instruments, using methods such as speckle \citep{horch14} or lucky imaging \citep{law06}, to be able to detect a large fraction of the sub-arcsecond companions which are not recovered by \textit{Gaia}. In addition, as the TESS targets will be significantly closer than for \textit{Kepler}, the on-sky angular separation of binaries will increase, allowing a larger fraction of binaries to be detected by diffraction-limited instruments on meter-class telescopes.

In addition, with multiple stars contributing to a single cumulative TESS light curve in which a purported planet transit signal is detected, it may be unclear which star is the source of the brightness dip (i.e., whether the transit is indeed a planet around the bright star, or a faint background eclipsing binary). The \textit{Kepler} pipeline identifies some astrophysical false positives through a variety of tests, such as significant secondary transit events or in- and out-of-transit centroid shifts \citep{coughlin15}. The latter of these tests will be more difficult with the lower resolution and coarser plate scale of TESS.

\section{Conclusion}
\label{sec:conclusion}

We found that the majority of binaries from the Robo-AO \textit{Kepler} survey with separations greater than 1\arcsec were recovered in \textit{Gaia} DR2 with magnitude contrasts as large as 7. Binaries with separations less than 0\farcs75 were typically not recovered, regardless of secondary brightness. We find that the recovery rate of binaries by \textit{Gaia} is not dependent on position angle. We found 177 nearby stars to \textit{Kepler} planetary candidates in DR2 that were not detected by Robo-AO. These newly detected stars are faint and likely not bound to the primary, and their impact on the planet candidate radii estimates is likely minimal. Between Robo-AO and \textit{Gaia}, we found that 18.7$\pm$0.7\% of \textit{Kepler} planet candidate hosts have nearby stars within 4\arcsec. In addition, we found 36 nearby stars around 35 K2 planetary candidates, and the K2 planet hosts displayed a significantly lower nearby star fraction rate than the \textit{Kepler} planet hosts. 

With years of observations to come, it is expected that the sensitivity of \textit{Gaia} will improve in later data releases, converging on the simulated recovery rate reported by \citet{arenou17}, with most binaries outside of 0\farcs5 detected. At present, \textit{Gaia} DR2 will improve initial TESS planet radius estimates by identifying contaminating sources within the same pixel as the planet host star. For precise characterization and confirmation, however, further ground-based high-resolution follow-up observations will be required.

A future analysis will use existing Keck-AO observations of \textit{Kepler} planet candidates performed by the Robo-AO team, as well as available archival data, to further test the sensitivity to close stellar binaries in \textit{Gaia} DR2 and subsequent catalogs. The astrometric and photometric precision achieved by \textit{Gaia} for stars in close proximity will be compared to that of single stars. With the deep imaging available with a large-aperture telescope, we will also be able to confirm or refute the existence of faint, potentially spurious, sources detected by \textit{Gaia} near bright stars.

\acknowledgments

This work uses data from research supported by the NASA Exoplanets Research Program, grant $\#$NNX 15AC91G. C.Z. is supported by a Dunlap Fellowship at the Dunlap Institute for Astronomy \& Astrophysics, funded through an endowment established by the Dunlap family and the University of Toronto. C.B. acknowledges support from the Alfred P. Sloan Foundation. T.M is supported by NASA grant $\#$NNX 14AE11G under the Kepler Participating Scientist Program. S.M. acknowledges support from the National Science Foundation award AST-1517592. D.H. acknowledges support by the National Science Foundation (AST-1717000) and the National Aeronautics and Space Administration under Grant NNX14AB92G issued through the Kepler Participating Scientist Program. The authors thank the Research Corporation for hosting the 2018 Time Domain Astronomy Scialog, where the idea for this project originated.

The Robo-AO team thanks NSF and NOAO for making the Kitt Peak 2.1-m telescope available. Robo-AO KP is a partnership between the California Institute of Technology, the University of Hawai‘i, the University of North Carolina at Chapel Hill, the Inter-University Centre for Astronomy and Astrophysics (IUCAA) at Pune, India, and the National Central University, Taiwan. The Murty family feels very happy to have added a small value to this important project. Robo-AO KP is also supported by grants from the John Templeton Foundation and the Mt. Cuba Astronomical Foundation. 

This research has made use of the NASA Exoplanet Archive, which is operated by the California Institute of Technology, under contract with the National Aeronautics and Space Administration under the Exoplanet Exploration Program.  This work has made use of data from the European Space Agency (ESA) mission \textit{Gaia} (https://www.cosmos.esa.int/gaia), processed by the Gaia Data Processing and Analysis Consortium (DPAC, https://www.cosmos.esa.int/web/gaia/dpac/consortium). Funding for the DPAC has been provided by national institutions, in particular the institutions participating in the \textit{Gaia} Multilateral Agreement. This work made use of the gaia-kepler.fun crossmatch database created by Megan Bedell.

We thank the anonymous referee for her or his careful analysis and useful comments on the manuscript.

{\it Facilities:} \facility{PO:1.5m (Robo-AO)}, \facility{KPNO:2.1m	(Robo-AO), \facility{Gaia}}

\bibliography{references}

\begin{thebibliography}{40}
\expandafter\ifx\csname natexlab\endcsname\relax\def\natexlab#1{#1}\fi

\bibitem[{{Andrae} {et~al.}(2018){Andrae}, {Fouesneau}, {Creevey}, {Ordenovic},
  {Mary}, {Burlacu}, {Chaoul}, {Jean-Antoine-Piccolo}, {Kordopatis}, {Korn},
  {Lebreton}, {Panem}, {Pichon}, {Thevenin}, {Walmsley}, \&
  {Bailer-Jones}}]{andrae18}
{Andrae}, R., {Fouesneau}, M., {Creevey}, O., {Ordenovic}, C., {Mary}, N.,
  {Burlacu}, A., {Chaoul}, L., {Jean-Antoine-Piccolo}, A., {Kordopatis}, G.,
  {Korn}, A., {Lebreton}, Y., {Panem}, C., {Pichon}, B., {Thevenin}, F.,
  {Walmsley}, G., \& {Bailer-Jones}, C.~A.~L. 2018, ArXiv e-prints

\bibitem[{{Arenou} {et~al.}(2018){Arenou}, {Luri}, {Babusiaux}, {Fabricius},
  {Helmi}, {Muraveva}, {Robin}, {Spoto}, {Vallenari}, {Antoja},
  {Cantat-Gaudin}, {Jordi}, {Leclerc}, {Reyl{\'e}}, {Romero-G{\'o}mez}, {Shih},
  {Soria}, {Barache}, {Bossini}, {Bragaglia}, {Breddels}, {Fabrizio},
  {Lambert}, {Marrese}, {Massari}, {Moitinho}, {Robichon}, {Ruiz-Dern},
  {Sordo}, {Veljanoski}, {Di Matteo}, {Eyer}, {Jasniewicz}, {Pancino},
  {Soubiran}, {Spagna}, {Tanga}, {Turon}, \& {Zurbach}}]{arenou18}
{Arenou}, F., {Luri}, X., {Babusiaux}, C., {Fabricius}, C., {Helmi}, A.,
  {Muraveva}, T., {Robin}, A.~C., {Spoto}, F., {Vallenari}, A., {Antoja}, T.,
  {Cantat-Gaudin}, T., {Jordi}, C., {Leclerc}, N., {Reyl{\'e}}, C.,
  {Romero-G{\'o}mez}, M., {Shih}, I., {Soria}, S., {Barache}, C., {Bossini},
  D., {Bragaglia}, A., {Breddels}, M.~A., {Fabrizio}, M., {Lambert}, S.,
  {Marrese}, P.~M., {Massari}, D., {Moitinho}, A., {Robichon}, N., {Ruiz-Dern},
  L., {Sordo}, R., {Veljanoski}, J., {Di Matteo}, P., {Eyer}, L., {Jasniewicz},
  G., {Pancino}, E., {Soubiran}, C., {Spagna}, A., {Tanga}, P., {Turon}, C., \&
  {Zurbach}, C. 2018, ArXiv e-prints

\bibitem[{{Arenou} {et~al.}(2017){Arenou}, {Luri}, {Babusiaux}, {Fabricius},
  {Helmi}, {Robin}, {Vallenari}, {Blanco-Cuaresma}, {Cantat-Gaudin},
  {Findeisen}, {Reyl{\'e}}, {Ruiz-Dern}, {Sordo}, {Turon}, {Walton}, {Shih},
  {Antiche}, {Barache}, {Barros}, {Breddels}, {Carrasco}, {Costigan},
  {Diakit{\'e}}, {Eyer}, {Figueras}, {Galluccio}, {Heu}, {Jordi},
  {Krone-Martins}, {Lallement}, {Lambert}, {Leclerc}, {Marrese}, {Moitinho},
  {Mor}, {Romero-G{\'o}mez}, {Sartoretti}, {Soria}, {Soubiran}, {Souchay},
  {Veljanoski}, {Ziaeepour}, {Giuffrida}, {Pancino}, \& {Bragaglia}}]{arenou17}
{Arenou}, F., {Luri}, X., {Babusiaux}, C., {Fabricius}, C., {Helmi}, A.,
  {Robin}, A.~C., {Vallenari}, A., {Blanco-Cuaresma}, S., {Cantat-Gaudin}, T.,
  {Findeisen}, K., {Reyl{\'e}}, C., {Ruiz-Dern}, L., {Sordo}, R., {Turon}, C.,
  {Walton}, N.~A., {Shih}, I.-C., {Antiche}, E., {Barache}, C., {Barros}, M.,
  {Breddels}, M., {Carrasco}, J.~M., {Costigan}, G., {Diakit{\'e}}, S., {Eyer},
  L., {Figueras}, F., {Galluccio}, L., {Heu}, J., {Jordi}, C., {Krone-Martins},
  A., {Lallement}, R., {Lambert}, S., {Leclerc}, N., {Marrese}, P.~M.,
  {Moitinho}, A., {Mor}, R., {Romero-G{\'o}mez}, M., {Sartoretti}, P., {Soria},
  S., {Soubiran}, C., {Souchay}, J., {Veljanoski}, J., {Ziaeepour}, H.,
  {Giuffrida}, G., {Pancino}, E., \& {Bragaglia}, A. 2017, \aap, 599, A50

\bibitem[{{Bailer-Jones} {et~al.}(2018){Bailer-Jones}, {Rybizki}, {Fouesneau},
  {Mantelet}, \& {Andrae}}]{bailerjones18}
{Bailer-Jones}, C.~A.~L., {Rybizki}, J., {Fouesneau}, M., {Mantelet}, G., \&
  {Andrae}, R. 2018, ArXiv e-prints

\bibitem[{{Baranec} {et~al.}(2017){Baranec}, {Riddle}, \& {Law}}]{baranec17}
{Baranec}, C., {Riddle}, R., \& {Law}, N.~M. 2017, ArXiv e-prints

\bibitem[{{Baranec} {et~al.}(2014){Baranec}, {Riddle}, {Law}, {Ramaprakash},
  {Tendulkar}, {Hogstrom}, {Bui}, {Burse}, {Chordia}, {Das}, {Dekany},
  {Kulkarni}, \& {Punnadi}}]{baranec14}
{Baranec}, C., {Riddle}, R., {Law}, N.~M., {Ramaprakash}, A.~N., {Tendulkar},
  S.~P., {Hogstrom}, K., {Bui}, K., {Burse}, M., {Chordia}, P., {Das}, H.,
  {Dekany}, R.~G., {Kulkarni}, S., \& {Punnadi}, S. 2014, \apjl, 790, L8

\bibitem[{{Baranec} {et~al.}(2016){Baranec}, {Ziegler}, {Law}, {Morton},
  {Riddle}, {Atkinson}, {Schonhut}, \& {Crepp}}]{baranec16}
{Baranec}, C., {Ziegler}, C., {Law}, N.~M., {Morton}, T., {Riddle}, R.,
  {Atkinson}, D., {Schonhut}, J., \& {Crepp}, J. 2016, \aj, 152, 18

\bibitem[{{Barclay} {et~al.}(2018){Barclay}, {Pepper}, \&
  {Quintana}}]{barclay18}
{Barclay}, T., {Pepper}, J., \& {Quintana}, E.~V. 2018, ArXiv e-prints

\bibitem[{{Berger} {et~al.}(2018){Berger}, {Huber}, {Gaidos}, \& {van
  Saders}}]{berger18}
{Berger}, T.~A., {Huber}, D., {Gaidos}, E., \& {van Saders}, J.~L. 2018, ArXiv
  e-prints

\bibitem[{{Chabrier} {et~al.}(2000){Chabrier}, {Baraffe}, {Allard}, \&
  {Hauschildt}}]{chabrier00}
{Chabrier}, G., {Baraffe}, I., {Allard}, F., \& {Hauschildt}, P. 2000, \apj,
  542, 464

\bibitem[{{Ciardi} {et~al.}(2015){Ciardi}, {Beichman}, {Horch}, \&
  {Howell}}]{ciardi15}
{Ciardi}, D.~R., {Beichman}, C.~A., {Horch}, E.~P., \& {Howell}, S.~B. 2015,
  \apj, 805, 16

\bibitem[{{Coughlin} {et~al.}(2016){Coughlin}, {Mullally}, {Thompson}, {Rowe},
  {Burke}, {Latham}, {Batalha}, {Ofir}, {Quarles}, {Henze}, {Wolfgang},
  {Caldwell}, {Bryson}, {Shporer}, {Catanzarite}, {Akeson}, {Barclay},
  {Borucki}, {Boyajian}, {Campbell}, {Christiansen}, {Girouard}, {Haas},
  {Howell}, {Huber}, {Jenkins}, {Li}, {Patil-Sabale}, {Quintana}, {Ramirez},
  {Seader}, {Smith}, {Tenenbaum}, {Twicken}, \& {Zamudio}}]{coughlin15}
{Coughlin}, J.~L., {Mullally}, F., {Thompson}, S.~E., {Rowe}, J.~F., {Burke},
  C.~J., {Latham}, D.~W., {Batalha}, N.~M., {Ofir}, A., {Quarles}, B.~L.,
  {Henze}, C.~E., {Wolfgang}, A., {Caldwell}, D.~A., {Bryson}, S.~T.,
  {Shporer}, A., {Catanzarite}, J., {Akeson}, R., {Barclay}, T., {Borucki},
  W.~J., {Boyajian}, T.~S., {Campbell}, J.~R., {Christiansen}, J.~L.,
  {Girouard}, F.~R., {Haas}, M.~R., {Howell}, S.~B., {Huber}, D., {Jenkins},
  J.~M., {Li}, J., {Patil-Sabale}, A., {Quintana}, E.~V., {Ramirez}, S.,
  {Seader}, S., {Smith}, J.~C., {Tenenbaum}, P., {Twicken}, J.~D., \&
  {Zamudio}, K.~A. 2016, \apjs, 224, 12

\bibitem[{{Crossfield} {et~al.}(2016){Crossfield}, {Ciardi}, {Petigura},
  {Sinukoff}, {Schlieder}, {Howard}, {Beichman}, {Isaacson}, {Dressing},
  {Christiansen}, {Fulton}, {L{\'e}pine}, {Weiss}, {Hirsch}, {Livingston},
  {Baranec}, {Law}, {Riddle}, {Ziegler}, {Howell}, {Horch}, {Everett}, {Teske},
  {Martinez}, {Obermeier}, {Benneke}, {Scott}, {Deacon}, {Aller}, {Hansen},
  {Mancini}, {Ciceri}, {Brahm}, {Jord{\'a}n}, {Knutson}, {Henning}, {Bonnefoy},
  {Liu}, {Crepp}, {Lothringer}, {Hinz}, {Bailey}, {Skemer}, \&
  {Defrere}}]{crossfield16}
{Crossfield}, I.~J.~M., {Ciardi}, D.~R., {Petigura}, E.~A., {Sinukoff}, E.,
  {Schlieder}, J.~E., {Howard}, A.~W., {Beichman}, C.~A., {Isaacson}, H.,
  {Dressing}, C.~D., {Christiansen}, J.~L., {Fulton}, B.~J., {L{\'e}pine}, S.,
  {Weiss}, L., {Hirsch}, L., {Livingston}, J., {Baranec}, C., {Law}, N.~M.,
  {Riddle}, R., {Ziegler}, C., {Howell}, S.~B., {Horch}, E., {Everett}, M.,
  {Teske}, J., {Martinez}, A.~O., {Obermeier}, C., {Benneke}, B., {Scott}, N.,
  {Deacon}, N., {Aller}, K.~M., {Hansen}, B.~M.~S., {Mancini}, L., {Ciceri},
  S., {Brahm}, R., {Jord{\'a}n}, A., {Knutson}, H.~A., {Henning}, T.,
  {Bonnefoy}, M., {Liu}, M.~C., {Crepp}, J.~R., {Lothringer}, J., {Hinz}, P.,
  {Bailey}, V., {Skemer}, A., \& {Defrere}, D. 2016, \apjs, 226, 7

\bibitem[{{de Bruijne} {et~al.}(2015){de Bruijne}, {Allen}, {Azaz},
  {Krone-Martins}, {Prod'homme}, \& {Hestroffer}}]{debruijne15}
{de Bruijne}, J.~H.~J., {Allen}, M., {Azaz}, S., {Krone-Martins}, A.,
  {Prod'homme}, T., \& {Hestroffer}, D. 2015, \aap, 576, A74

\bibitem[{{Dotter} {et~al.}(2008){Dotter}, {Chaboyer}, {Jevremovi{\'c}},
  {Kostov}, {Baron}, \& {Ferguson}}]{dotter08}
{Dotter}, A., {Chaboyer}, B., {Jevremovi{\'c}}, D., {Kostov}, V., {Baron}, E.,
  \& {Ferguson}, J.~W. 2008, \apjs, 178, 89

\bibitem[{{Fabricius} {et~al.}(2016){Fabricius}, {Bastian}, {Portell},
  {Casta{\~n}eda}, {Davidson}, {Hambly}, {Clotet}, {Biermann}, {Mora},
  {Busonero}, {Riva}, {Brown}, {Smart}, {Lammers}, {Torra}, {Drimmel},
  {Gracia}, {L{\"o}ffler}, {Spagna}, {Lindegren}, {Klioner}, {Andrei}, {Bach},
  {Bramante}, {Br{\"u}semeister}, {Busso}, {Carrasco}, {Gai}, {Garralda},
  {Gonz{\'a}lez-Vidal}, {Guerra}, {Hauser}, {Jordan}, {Jordi}, {Lenhardt},
  {Mignard}, {Messineo}, {Mulone}, {Serraller}, {Stampa}, {Tanga}, {van
  Elteren}, {van Reeven}, {Voss}, {Abbas}, {Allasia}, {Altmann}, {Anton},
  {Barache}, {Becciani}, {Berthier}, {Bianchi}, {Bombrun}, {Bouquillon},
  {Bourda}, {Bucciarelli}, {Butkevich}, {Buzzi}, {Cancelliere}, {Carlucci},
  {Charlot}, {Collins}, {Comoretto}, {Cross}, {Crosta}, {de Felice}, {Fienga},
  {Figueras}, {Fraile}, {Geyer}, {Hernandez}, {Hobbs}, {Hofmann}, {Liao},
  {Licata}, {Martino}, {McMillan}, {Michalik}, {Morbidelli}, {Parsons},
  {Pecoraro}, {Ramos-Lerate}, {Sarasso}, {Siddiqui}, {Steele},
  {Steidelm{\"u}ller}, {Taris}, {Vecchiato}, {Abreu}, {Anglada}, {Boudreault},
  {Cropper}, {Holl}, {Cheek}, {Crowley}, {Fleitas}, {Hutton}, {Osinde},
  {Rowell}, {Salguero}, {Utrilla}, {Blagorodnova}, {Soffel}, {Osorio},
  {Vicente}, {Cambras}, \& {Bernstein}}]{dr1_sourcelist}
{Fabricius}, C., {Bastian}, U., {Portell}, J., {Casta{\~n}eda}, J., {Davidson},
  M., {Hambly}, N.~C., {Clotet}, M., {Biermann}, M., {Mora}, A., {Busonero},
  D., {Riva}, A., {Brown}, A.~G.~A., {Smart}, R., {Lammers}, U., {Torra}, J.,
  {Drimmel}, R., {Gracia}, G., {L{\"o}ffler}, W., {Spagna}, A., {Lindegren},
  L., {Klioner}, S., {Andrei}, A., {Bach}, N., {Bramante}, L.,
  {Br{\"u}semeister}, T., {Busso}, G., {Carrasco}, J.~M., {Gai}, M.,
  {Garralda}, N., {Gonz{\'a}lez-Vidal}, J.~J., {Guerra}, R., {Hauser}, M.,
  {Jordan}, S., {Jordi}, C., {Lenhardt}, H., {Mignard}, F., {Messineo}, R.,
  {Mulone}, A., {Serraller}, I., {Stampa}, U., {Tanga}, P., {van Elteren}, A.,
  {van Reeven}, W., {Voss}, H., {Abbas}, U., {Allasia}, W., {Altmann}, M.,
  {Anton}, S., {Barache}, C., {Becciani}, U., {Berthier}, J., {Bianchi}, L.,
  {Bombrun}, A., {Bouquillon}, S., {Bourda}, G., {Bucciarelli}, B.,
  {Butkevich}, A., {Buzzi}, R., {Cancelliere}, R., {Carlucci}, T., {Charlot},
  P., {Collins}, R., {Comoretto}, G., {Cross}, N., {Crosta}, M., {de Felice},
  F., {Fienga}, A., {Figueras}, F., {Fraile}, E., {Geyer}, R., {Hernandez}, J.,
  {Hobbs}, D., {Hofmann}, W., {Liao}, S., {Licata}, E., {Martino}, M.,
  {McMillan}, P.~J., {Michalik}, D., {Morbidelli}, R., {Parsons}, P.,
  {Pecoraro}, M., {Ramos-Lerate}, M., {Sarasso}, M., {Siddiqui}, H., {Steele},
  I., {Steidelm{\"u}ller}, H., {Taris}, F., {Vecchiato}, A., {Abreu}, A.,
  {Anglada}, E., {Boudreault}, S., {Cropper}, M., {Holl}, B., {Cheek}, N.,
  {Crowley}, C., {Fleitas}, J.~M., {Hutton}, A., {Osinde}, J., {Rowell}, N.,
  {Salguero}, E., {Utrilla}, E., {Blagorodnova}, N., {Soffel}, M., {Osorio},
  J., {Vicente}, D., {Cambras}, J., \& {Bernstein}, H.~H. 2016, \aap, 595, A3

\bibitem[{{Gaia Collaboration} {et~al.}(2018{\natexlab{a}}){Gaia
  Collaboration}, {Brown}, {Vallenari}, {Prusti}, {de Bruijne}, {Babusiaux}, \&
  {Bailer-Jones}}]{gaiadr2}
{Gaia Collaboration}, {Brown}, A.~G.~A., {Vallenari}, A., {Prusti}, T., {de
  Bruijne}, J.~H.~J., {Babusiaux}, C., \& {Bailer-Jones}, C.~A.~L.
  2018{\natexlab{a}}, ArXiv e-prints

\bibitem[{{Gaia Collaboration} {et~al.}(2016){Gaia Collaboration}, {Prusti},
  {de Bruijne}, {Brown}, {Vallenari}, {Babusiaux}, {Bailer-Jones}, {Bastian},
  {Biermann}, {Evans}, \& et~al.}]{gaia2016b}
{Gaia Collaboration}, {Prusti}, T., {de Bruijne}, J.~H.~J., {Brown}, A.~G.~A.,
  {Vallenari}, A., {Babusiaux}, C., {Bailer-Jones}, C.~A.~L., {Bastian}, U.,
  {Biermann}, M., {Evans}, D.~W., \& et~al. 2016, \aap, 595, A1

\bibitem[{{Gaia Collaboration} {et~al.}(2018{\natexlab{b}}){Gaia
  Collaboration}, {Spoto}, {Tanga}, {Mignard}, {Berthier}, {Carry}, \&
  {Cellino}}]{sso_gaia}
{Gaia Collaboration}, {Spoto}, F., {Tanga}, P., {Mignard}, F., {Berthier}, J.,
  {Carry}, B., \& {Cellino}, A. 2018{\natexlab{b}}, ArXiv e-prints

\bibitem[{{Horch} {et~al.}(2014){Horch}, {Howell}, {Everett}, \&
  {Ciardi}}]{horch14}
{Horch}, E.~P., {Howell}, S.~B., {Everett}, M.~E., \& {Ciardi}, D.~R. 2014,
  \apj, 795, 60

\bibitem[{{Howard} {et~al.}(2012){Howard}, {Marcy}, {Bryson}, {Jenkins},
  {Rowe}, {Batalha}, {Borucki}, {Koch}, {Dunham}, {Gautier}, {Van Cleve},
  {Cochran}, {Latham}, {Lissauer}, {Torres}, {Brown}, {Gilliland}, {Buchhave},
  {Caldwell}, {Christensen-Dalsgaard}, {Ciardi}, {Fressin}, {Haas}, {Howell},
  {Kjeldsen}, {Seager}, {Rogers}, {Sasselov}, {Steffen}, {Basri},
  {Charbonneau}, {Christiansen}, {Clarke}, {Dupree}, {Fabrycky}, {Fischer},
  {Ford}, {Fortney}, {Tarter}, {Girouard}, {Holman}, {Johnson}, {Klaus},
  {Machalek}, {Moorhead}, {Morehead}, {Ragozzine}, {Tenenbaum}, {Twicken},
  {Quinn}, {Isaacson}, {Shporer}, {Lucas}, {Walkowicz}, {Welsh}, {Boss},
  {Devore}, {Gould}, {Smith}, {Morris}, {Prsa}, {Morton}, {Still}, {Thompson},
  {Mullally}, {Endl}, \& {MacQueen}}]{howard12}
{Howard}, A.~W., {Marcy}, G.~W., {Bryson}, S.~T., {Jenkins}, J.~M., {Rowe},
  J.~F., {Batalha}, N.~M., {Borucki}, W.~J., {Koch}, D.~G., {Dunham}, E.~W.,
  {Gautier}, III, T.~N., {Van Cleve}, J., {Cochran}, W.~D., {Latham}, D.~W.,
  {Lissauer}, J.~J., {Torres}, G., {Brown}, T.~M., {Gilliland}, R.~L.,
  {Buchhave}, L.~A., {Caldwell}, D.~A., {Christensen-Dalsgaard}, J., {Ciardi},
  D., {Fressin}, F., {Haas}, M.~R., {Howell}, S.~B., {Kjeldsen}, H., {Seager},
  S., {Rogers}, L., {Sasselov}, D.~D., {Steffen}, J.~H., {Basri}, G.~S.,
  {Charbonneau}, D., {Christiansen}, J., {Clarke}, B., {Dupree}, A.,
  {Fabrycky}, D.~C., {Fischer}, D.~A., {Ford}, E.~B., {Fortney}, J.~J.,
  {Tarter}, J., {Girouard}, F.~R., {Holman}, M.~J., {Johnson}, J.~A., {Klaus},
  T.~C., {Machalek}, P., {Moorhead}, A.~V., {Morehead}, R.~C., {Ragozzine}, D.,
  {Tenenbaum}, P., {Twicken}, J.~D., {Quinn}, S.~N., {Isaacson}, H., {Shporer},
  A., {Lucas}, P.~W., {Walkowicz}, L.~M., {Welsh}, W.~F., {Boss}, A., {Devore},
  E., {Gould}, A., {Smith}, J.~C., {Morris}, R.~L., {Prsa}, A., {Morton},
  T.~D., {Still}, M., {Thompson}, S.~E., {Mullally}, F., {Endl}, M., \&
  {MacQueen}, P.~J. 2012, \apjs, 201, 15

\bibitem[{{Huber} {et~al.}(2016){Huber}, {Bryson}, {Haas}, {Barclay},
  {Barentsen}, {Howell}, {Sharma}, {Stello}, \& {Thompson}}]{huber16}
{Huber}, D., {Bryson}, S.~T., {Haas}, M.~R., {Barclay}, T., {Barentsen}, G.,
  {Howell}, S.~B., {Sharma}, S., {Stello}, D., \& {Thompson}, S.~E. 2016,
  \apjs, 224, 2

\bibitem[{{Janson} {et~al.}(2012){Janson}, {Hormuth}, {Bergfors}, {Brandner},
  {Hippler}, {Daemgen}, {Kudryavtseva}, {Schmalzl}, {Schnupp}, \&
  {Henning}}]{janson12}
{Janson}, M., {Hormuth}, F., {Bergfors}, C., {Brandner}, W., {Hippler}, S.,
  {Daemgen}, S., {Kudryavtseva}, N., {Schmalzl}, E., {Schnupp}, C., \&
  {Henning}, T. 2012, \apj, 754, 44

\bibitem[{{Jensen-Clem} {et~al.}(2018){Jensen-Clem}, {Duev}, {Riddle},
  {Salama}, {Baranec}, {Law}, {Kulkarni}, \& {Ramprakash}}]{jc18}
{Jensen-Clem}, R., {Duev}, D.~A., {Riddle}, R., {Salama}, M., {Baranec}, C.,
  {Law}, N.~M., {Kulkarni}, S.~R., \& {Ramprakash}, A.~N. 2018, \aj, 155, 32

\bibitem[{{Kraus} {et~al.}(2016){Kraus}, {Ireland}, {Huber}, {Mann}, \&
  {Dupuy}}]{kraus16}
{Kraus}, A.~L., {Ireland}, M.~J., {Huber}, D., {Mann}, A.~W., \& {Dupuy}, T.~J.
  2016, \aj, 152, 8

\bibitem[{{Law} {et~al.}(2006){Law}, {Hodgkin}, \& {Mackay}}]{law06}
{Law}, N.~M., {Hodgkin}, S.~T., \& {Mackay}, C.~D. 2006, \mnras, 368, 1917

\bibitem[{{Law} {et~al.}(2005){Law}, {Hodgkin}, {Mackay}, \& {Baldwin}}]{law05}
{Law}, N.~M., {Hodgkin}, S.~T., {Mackay}, C.~D., \& {Baldwin}, J.~E. 2005,
  Astronomische Nachrichten, 326, 1024

\bibitem[{{Law} {et~al.}(2014){Law}, {Morton}, {Baranec}, {Riddle},
  {Ravichandran}, {Ziegler}, {Johnson}, {Tendulkar}, {Bui}, {Burse}, {Das},
  {Dekany}, {Kulkarni}, {Punnadi}, \& {Ramaprakash}}]{law14}
{Law}, N.~M., {Morton}, T., {Baranec}, C., {Riddle}, R., {Ravichandran}, G.,
  {Ziegler}, C., {Johnson}, J.~A., {Tendulkar}, S.~P., {Bui}, K., {Burse},
  M.~P., {Das}, H.~K., {Dekany}, R.~G., {Kulkarni}, S., {Punnadi}, S., \&
  {Ramaprakash}, A.~N. 2014, \apj, 791, 35

\bibitem[{{Mathur} {et~al.}(2017){Mathur}, {Huber}, {Batalha}, {Ciardi},
  {Bastien}, {Bieryla}, {Buchhave}, {Cochran}, {Endl}, {Esquerdo}, {Furlan},
  {Howard}, {Howell}, {Isaacson}, {Latham}, {MacQueen}, \& {Silva}}]{dr25}
{Mathur}, S., {Huber}, D., {Batalha}, N.~M., {Ciardi}, D.~R., {Bastien}, F.~A.,
  {Bieryla}, A., {Buchhave}, L.~A., {Cochran}, W.~D., {Endl}, M., {Esquerdo},
  G.~A., {Furlan}, E., {Howard}, A., {Howell}, S.~B., {Isaacson}, H., {Latham},
  D.~W., {MacQueen}, P.~J., \& {Silva}, D.~R. 2017, \apjs, 229, 30

\bibitem[{{Moe} \& {Di Stefano}(2017)}]{moe17}
{Moe}, M. \& {Di Stefano}, R. 2017, \apjs, 230, 15

\bibitem[{{Raghavan} {et~al.}(2010){Raghavan}, {McAlister}, {Henry}, {Latham},
  {Marcy}, {Mason}, {Gies}, {White}, \& {ten Brummelaar}}]{raghavan10}
{Raghavan}, D., {McAlister}, H.~A., {Henry}, T.~J., {Latham}, D.~W., {Marcy},
  G.~W., {Mason}, B.~D., {Gies}, D.~R., {White}, R.~J., \& {ten Brummelaar},
  T.~A. 2010, \apjs, 190, 1

\bibitem[{{Ricker} {et~al.}(2014){Ricker}, {Winn}, {Vanderspek}, {Latham},
  {Bakos}, {Bean}, {Berta-Thompson}, {Brown}, {Buchhave}, {Butler}, {Butler},
  {Chaplin}, {Charbonneau}, {Christensen-Dalsgaard}, {Clampin}, {Deming},
  {Doty}, {De Lee}, {Dressing}, {Dunham}, {Endl}, {Fressin}, {Ge}, {Henning},
  {Holman}, {Howard}, {Ida}, {Jenkins}, {Jernigan}, {Johnson}, {Kaltenegger},
  {Kawai}, {Kjeldsen}, {Laughlin}, {Levine}, {Lin}, {Lissauer}, {MacQueen},
  {Marcy}, {McCullough}, {Morton}, {Narita}, {Paegert}, {Palle}, {Pepe},
  {Pepper}, {Quirrenbach}, {Rinehart}, {Sasselov}, {Sato}, {Seager},
  {Sozzetti}, {Stassun}, {Sullivan}, {Szentgyorgyi}, {Torres}, {Udry}, \&
  {Villasenor}}]{ricker15}
{Ricker}, G.~R., {Winn}, J.~N., {Vanderspek}, R., {Latham}, D.~W., {Bakos},
  G.~{\'A}., {Bean}, J.~L., {Berta-Thompson}, Z.~K., {Brown}, T.~M.,
  {Buchhave}, L., {Butler}, N.~R., {Butler}, R.~P., {Chaplin}, W.~J.,
  {Charbonneau}, D., {Christensen-Dalsgaard}, J., {Clampin}, M., {Deming}, D.,
  {Doty}, J., {De Lee}, N., {Dressing}, C., {Dunham}, E.~W., {Endl}, M.,
  {Fressin}, F., {Ge}, J., {Henning}, T., {Holman}, M.~J., {Howard}, A.~W.,
  {Ida}, S., {Jenkins}, J., {Jernigan}, G., {Johnson}, J.~A., {Kaltenegger},
  L., {Kawai}, N., {Kjeldsen}, H., {Laughlin}, G., {Levine}, A.~M., {Lin}, D.,
  {Lissauer}, J.~J., {MacQueen}, P., {Marcy}, G., {McCullough}, P.~R.,
  {Morton}, T.~D., {Narita}, N., {Paegert}, M., {Palle}, E., {Pepe}, F.,
  {Pepper}, J., {Quirrenbach}, A., {Rinehart}, S.~A., {Sasselov}, D., {Sato},
  B., {Seager}, S., {Sozzetti}, A., {Stassun}, K.~G., {Sullivan}, P.,
  {Szentgyorgyi}, A., {Torres}, G., {Udry}, S., \& {Villasenor}, J. 2014, in
  \procspie, Vol. 9143, Space Telescopes and Instrumentation 2014: Optical,
  Infrared, and Millimeter Wave, 914320

\bibitem[{{The Astropy Collaboration} {et~al.}(2018){The Astropy
  Collaboration}, {Price-Whelan}, {Sip{\H o}cz}, {G{\"u}nther}, {Lim},
  {Crawford}, {Conseil}, {Shupe}, {Craig}, {Dencheva}, {Ginsburg},
  {VanderPlas}, {Bradley}, {P{\'e}rez-Su{\'a}rez}, {de Val-Borro}, {Aldcroft},
  {Cruz}, {Robitaille}, {Tollerud}, {Ardelean}, {Babej}, {Bachetti}, {Bakanov},
  {Bamford}, {Barentsen}, {Barmby}, {Baumbach}, {Berry}, {Biscani}, {Boquien},
  {Bostroem}, {Bouma}, {Brammer}, {Bray}, {Breytenbach}, {Buddelmeijer},
  {Burke}, {Calderone}, {Cano Rodr{\'{\i}}guez}, {Cara}, {Cardoso},
  {Cheedella}, {Copin}, {Crichton}, {D{\'A}vella}, {Deil}, {Depagne},
  {Dietrich}, {Donath}, {Droettboom}, {Earl}, {Erben}, {Fabbro}, {Ferreira},
  {Finethy}, {Fox}, {Garrison}, {Gibbons}, {Goldstein}, {Gommers}, {Greco},
  {Greenfield}, {Groener}, {Grollier}, {Hagen}, {Hirst}, {Homeier}, {Horton},
  {Hosseinzadeh}, {Hu}, {Hunkeler}, {Ivezi{\'c}}, {Jain}, {Jenness}, {Kanarek},
  {Kendrew}, {Kern}, {Kerzendorf}, {Khvalko}, {King}, {Kirkby}, {Kulkarni},
  {Kumar}, {Lee}, {Lenz}, {Littlefair}, {Ma}, {Macleod}, {Mastropietro},
  {McCully}, {Montagnac}, {Morris}, {Mueller}, {Mumford}, {Muna}, {Murphy},
  {Nelson}, {Nguyen}, {Ninan}, {N{\"o}the}, {Ogaz}, {Oh}, {Parejko}, {Parley},
  {Pascual}, {Patil}, {Patil}, {Plunkett}, {Prochaska}, {Rastogi}, {Reddy
  Janga}, {Sabater}, {Sakurikar}, {Seifert}, {Sherbert}, {Sherwood-Taylor},
  {Shih}, {Sick}, {Silbiger}, {Singanamalla}, {Singer}, {Sladen}, {Sooley},
  {Sornarajah}, {Streicher}, {Teuben}, {Thomas}, {Tremblay}, {Turner},
  {Terr{\'o}n}, {van Kerkwijk}, {de la Vega}, {Watkins}, {Weaver}, {Whitmore},
  {Woillez}, \& {Zabalza}}]{astropy}
{The Astropy Collaboration}, {Price-Whelan}, A.~M., {Sip{\H o}cz}, B.~M.,
  {G{\"u}nther}, H.~M., {Lim}, P.~L., {Crawford}, S.~M., {Conseil}, S.,
  {Shupe}, D.~L., {Craig}, M.~W., {Dencheva}, N., {Ginsburg}, A., {VanderPlas},
  J.~T., {Bradley}, L.~D., {P{\'e}rez-Su{\'a}rez}, D., {de Val-Borro}, M.,
  {Aldcroft}, T.~L., {Cruz}, K.~L., {Robitaille}, T.~P., {Tollerud}, E.~J.,
  {Ardelean}, C., {Babej}, T., {Bachetti}, M., {Bakanov}, A.~V., {Bamford},
  S.~P., {Barentsen}, G., {Barmby}, P., {Baumbach}, A., {Berry}, K.~L.,
  {Biscani}, F., {Boquien}, M., {Bostroem}, K.~A., {Bouma}, L.~G., {Brammer},
  G.~B., {Bray}, E.~M., {Breytenbach}, H., {Buddelmeijer}, H., {Burke}, D.~J.,
  {Calderone}, G., {Cano Rodr{\'{\i}}guez}, J.~L., {Cara}, M., {Cardoso},
  J.~V.~M., {Cheedella}, S., {Copin}, Y., {Crichton}, D., {D{\'A}vella}, D.,
  {Deil}, C., {Depagne}, {\'E}., {Dietrich}, J.~P., {Donath}, A., {Droettboom},
  M., {Earl}, N., {Erben}, T., {Fabbro}, S., {Ferreira}, L.~A., {Finethy}, T.,
  {Fox}, R.~T., {Garrison}, L.~H., {Gibbons}, S.~L.~J., {Goldstein}, D.~A.,
  {Gommers}, R., {Greco}, J.~P., {Greenfield}, P., {Groener}, A.~M.,
  {Grollier}, F., {Hagen}, A., {Hirst}, P., {Homeier}, D., {Horton}, A.~J.,
  {Hosseinzadeh}, G., {Hu}, L., {Hunkeler}, J.~S., {Ivezi{\'c}}, {\v Z}.,
  {Jain}, A., {Jenness}, T., {Kanarek}, G., {Kendrew}, S., {Kern}, N.~S.,
  {Kerzendorf}, W.~E., {Khvalko}, A., {King}, J., {Kirkby}, D., {Kulkarni},
  A.~M., {Kumar}, A., {Lee}, A., {Lenz}, D., {Littlefair}, S.~P., {Ma}, Z.,
  {Macleod}, D.~M., {Mastropietro}, M., {McCully}, C., {Montagnac}, S.,
  {Morris}, B.~M., {Mueller}, M., {Mumford}, S.~J., {Muna}, D., {Murphy},
  N.~A., {Nelson}, S., {Nguyen}, G.~H., {Ninan}, J.~P., {N{\"o}the}, M.,
  {Ogaz}, S., {Oh}, S., {Parejko}, J.~K., {Parley}, N., {Pascual}, S., {Patil},
  R., {Patil}, A.~A., {Plunkett}, A.~L., {Prochaska}, J.~X., {Rastogi}, T.,
  {Reddy Janga}, V., {Sabater}, J., {Sakurikar}, P., {Seifert}, M., {Sherbert},
  L.~E., {Sherwood-Taylor}, H., {Shih}, A.~Y., {Sick}, J., {Silbiger}, M.~T.,
  {Singanamalla}, S., {Singer}, L.~P., {Sladen}, P.~H., {Sooley}, K.~A.,
  {Sornarajah}, S., {Streicher}, O., {Teuben}, P., {Thomas}, S.~W., {Tremblay},
  G.~R., {Turner}, J.~E.~H., {Terr{\'o}n}, V., {van Kerkwijk}, M.~H., {de la
  Vega}, A., {Watkins}, L.~L., {Weaver}, B.~A., {Whitmore}, J.~B., {Woillez},
  J., \& {Zabalza}, V. 2018, ArXiv e-prints

\bibitem[{{Thompson} {et~al.}(2018){Thompson}, {Coughlin}, {Hoffman},
  {Mullally}, {Christiansen}, {Burke}, {Bryson}, {Batalha}, {Haas},
  {Catanzarite}, {Rowe}, {Barentsen}, {Caldwell}, {Clarke}, {Jenkins}, {Li},
  {Latham}, {Lissauer}, {Mathur}, {Morris}, {Seader}, {Smith}, {Klaus},
  {Twicken}, {Van Cleve}, {Wohler}, {Akeson}, {Ciardi}, {Cochran}, {Henze},
  {Howell}, {Huber}, {Pr{\v s}a}, {Ram{\'{\i}}rez}, {Morton}, {Barclay},
  {Campbell}, {Chaplin}, {Charbonneau}, {Christensen-Dalsgaard}, {Dotson},
  {Doyle}, {Dunham}, {Dupree}, {Ford}, {Geary}, {Girouard}, {Isaacson},
  {Kjeldsen}, {Quintana}, {Ragozzine}, {Shabram}, {Shporer}, {Silva Aguirre},
  {Steffen}, {Still}, {Tenenbaum}, {Welsh}, {Wolfgang}, {Zamudio}, {Koch}, \&
  {Borucki}}]{thompson18}
{Thompson}, S.~E., {Coughlin}, J.~L., {Hoffman}, K., {Mullally}, F.,
  {Christiansen}, J.~L., {Burke}, C.~J., {Bryson}, S., {Batalha}, N., {Haas},
  M.~R., {Catanzarite}, J., {Rowe}, J.~F., {Barentsen}, G., {Caldwell}, D.~A.,
  {Clarke}, B.~D., {Jenkins}, J.~M., {Li}, J., {Latham}, D.~W., {Lissauer},
  J.~J., {Mathur}, S., {Morris}, R.~L., {Seader}, S.~E., {Smith}, J.~C.,
  {Klaus}, T.~C., {Twicken}, J.~D., {Van Cleve}, J.~E., {Wohler}, B., {Akeson},
  R., {Ciardi}, D.~R., {Cochran}, W.~D., {Henze}, C.~E., {Howell}, S.~B.,
  {Huber}, D., {Pr{\v s}a}, A., {Ram{\'{\i}}rez}, S.~V., {Morton}, T.~D.,
  {Barclay}, T., {Campbell}, J.~R., {Chaplin}, W.~J., {Charbonneau}, D.,
  {Christensen-Dalsgaard}, J., {Dotson}, J.~L., {Doyle}, L., {Dunham}, E.~W.,
  {Dupree}, A.~K., {Ford}, E.~B., {Geary}, J.~C., {Girouard}, F.~R.,
  {Isaacson}, H., {Kjeldsen}, H., {Quintana}, E.~V., {Ragozzine}, D.,
  {Shabram}, M., {Shporer}, A., {Silva Aguirre}, V., {Steffen}, J.~H., {Still},
  M., {Tenenbaum}, P., {Welsh}, W.~F., {Wolfgang}, A., {Zamudio}, K.~A.,
  {Koch}, D.~G., \& {Borucki}, W.~J. 2018, \apjs, 235, 38

\bibitem[{{Zhang} {et~al.}(2013){Zhang}, {Pinfield}, {Burningham}, {Jones},
  {G{\'a}lvez-Ortiz}, {Catal{\'a}n}, {Smart}, {L{\'e}pine}, {Clarke},
  {Pavlenko}, {Murray}, {Kuznetsov}, {Day-Jones}, {Gomes}, {Marocco}, \&
  {Sip{\H o}cz}}]{zhang13}
{Zhang}, Z.~H., {Pinfield}, D.~J., {Burningham}, B., {Jones}, H.~R.~A.,
  {G{\'a}lvez-Ortiz}, M.~C., {Catal{\'a}n}, S., {Smart}, R.~L., {L{\'e}pine},
  S., {Clarke}, J.~R.~A., {Pavlenko}, Y.~V., {Murray}, D.~N., {Kuznetsov},
  M.~K., {Day-Jones}, A.~C., {Gomes}, J., {Marocco}, F., \& {Sip{\H o}cz}, B.
  2013, \mnras, 434, 1005

\bibitem[{{Ziegler} {et~al.}(2018{\natexlab{a}}){Ziegler}, {Law}, {Baranec},
  {Howard}, {Morton}, {Riddle}, {Duev}, {Salama}, {Jensen-Clem}, \&
  {Kulkarni}}]{ziegler18b}
{Ziegler}, C., {Law}, N.~M., {Baranec}, C., {Howard}, W., {Morton}, T.,
  {Riddle}, R., {Duev}, D.~A., {Salama}, M., {Jensen-Clem}, R., \& {Kulkarni},
  S.~R. 2018{\natexlab{a}}, ArXiv e-prints

\bibitem[{{Ziegler} {et~al.}(2018{\natexlab{b}}){Ziegler}, {Law}, {Baranec},
  {Riddle}, {Duev}, {Howard}, {Jensen-Clem}, {Kulkarni}, {Morton}, \&
  {Salama}}]{ziegler18}
{Ziegler}, C., {Law}, N.~M., {Baranec}, C., {Riddle}, R., {Duev}, D.~A.,
  {Howard}, W., {Jensen-Clem}, R., {Kulkarni}, S.~R., {Morton}, T., \&
  {Salama}, M. 2018{\natexlab{b}}, \aj, 155, 161

\bibitem[{{Ziegler} {et~al.}(2017{\natexlab{a}}){Ziegler}, {Law}, {Baranec},
  {Riddle}, {Duev}, {Howard}, {Jensen-Clem}, {Kulkarni}, \&
  {Salama}}]{ziegler18a}
{Ziegler}, C., {Law}, N.~M., {Baranec}, C., {Riddle}, R., {Duev}, D.~A.,
  {Howard}, W., {Jensen-Clem}, R., {Kulkarni}, S.~R., \& {Salama}, M.
  2017{\natexlab{a}}, ArXiv e-prints

\bibitem[{{Ziegler} {et~al.}(2015){Ziegler}, {Law}, {Baranec}, {Riddle}, \&
  {Fuchs}}]{ziegler15}
{Ziegler}, C., {Law}, N.~M., {Baranec}, C., {Riddle}, R.~L., \& {Fuchs}, J.~T.
  2015, \apj, 804, 30

\bibitem[{{Ziegler} {et~al.}(2017{\natexlab{b}}){Ziegler}, {Law}, {Morton},
  {Baranec}, {Riddle}, {Atkinson}, {Baker}, {Roberts}, \& {Ciardi}}]{ziegler16}
{Ziegler}, C., {Law}, N.~M., {Morton}, T., {Baranec}, C., {Riddle}, R.,
  {Atkinson}, D., {Baker}, A., {Roberts}, S., \& {Ciardi}, D.~R.
  2017{\natexlab{b}}, \aj, 153, 66

\end{thebibliography}

\LongTables
\clearpage
\tabletypesize{\scriptsize}

\begin{deluxetable*}{cccccccc}
\tablewidth{0pt}
\tablecaption{\label{tab:gaiadetections}Robo-AO Detected Nearby Stars to \textit{Kepler} Planetary Candidates in \textit{Gaia} DR2}
\tablehead{
\colhead{KOI} & \colhead{Sep.} & \colhead{P.A.} & \colhead{$\Delta$m}  & \colhead{Detection} & \colhead{Primary \textit{Gaia}} & \colhead{Secondary \textit{Gaia}}  & \colhead{Recovered in}\\
\colhead{} & \colhead{(\arcsec)} & \colhead{(deg.)} & \colhead{(mags)} & \colhead{reference} & \colhead{DR2 source ID} & \colhead{DR2 source ID} & \colhead{\textit{Gaia} DR2?}
}
\startdata

1 & 1.13 & 135 & 3.95 & L14 & 2131314401800665344 & \nodata & not-recovered \\ 
4 & 3.42 & 75 & 4.46 & B16 & 2052194374009877376 & 2052194374009877632 & recovered \\ 
13 & 1.16 & 279 & 0.19 & L14 & 2130632159134827392 & 2130632159130638464 & recovered \\ 
42 & 1.74 & 35 & 3.04 & B16 & 2106904148451706752 & 2106904148449360000 & recovered \\ 
44 & 3.42 & 123 & 4.03 & Z18 & 2082166236346124544 & 2082166236337520896 & recovered \\ 
51 & 3.51 & 161 & 2.63 & Z17 & 2077023717382584576 & \nodata & not-recovered \\ 
70 & 3.86 & 51 & 5.74 & Z18 & 2102548708017562112 & 2102548708017562368 & recovered \\ 
75 & 3.53 & 124 & 6.6 & Z18 & 2125795166969907968 & 2125795166970655104 & recovered \\
\enddata
\tablecomments{
\tablenotetext{}{Reference for nearby star detection: \citep[L14,][]{law14}, \citep[B16,][]{baranec16}, \citep[Z17,][]{ziegler16}, \citep[Z18,][]{ziegler18}. Provenance of reported companion properties is the detection reference publication.}
}
\end{deluxetable*}

\LongTables
\clearpage
\tabletypesize{\scriptsize}

\begin{deluxetable*}{cccccc}
\tablewidth{0pt}
\tablecaption{\label{tab:newgaiadetections}New Nearby Stars to \textit{Kepler} Planetary Candidates in \textit{Gaia} DR2}
\tablehead{
\colhead{KOI} & \colhead{Sep.} & \colhead{P.A.} & \colhead{$\Delta$m$_{G}$}  & \colhead{Primary \textit{Gaia}} & \colhead{Secondary \textit{Gaia}} \\
\colhead{} & \colhead{(\arcsec)} & \colhead{(deg.)} & \colhead{(mags)} & \colhead{DR2 source ID} & \colhead{DR2 source ID}
}
\startdata

18 & 3.48 & 111 & 7.27 & 2079018300195390464 & 2076015877539239680 \\ 
118 & 1.46 & 212 & 4.34 & 2099605968225288960 & 2099605968225289088 \\ 
217 & 3.9 & 326 & 5.16 & 2080095679848047872 & 2080095684146710400 \\ 
221 & 1.57 & 276 & 5.28 & 2100073393808483328 & 2100073398103782272 \\ 
235 & 3.54 & 13 & 5.64 & 2078125359313871488 & 2078125359309472768 \\ 
266 & 3.74 & 325 & 8.16 & 2078008703696509056 & 2078008707991780352 \\ 
344 & 3.57 & 210 & 6.87 & 2131736137528127744 & 2131736137526502656 \\ 
354 & 3.77 & 209 & 6.59 & 2134886100904948608 & 2134886066539848832 \\
\enddata

\tablecomments{
\tablenotetext{}{Provenance of reported companion properties is from \textit{Gaia} DR2 \citep{gaiadr2}.}
}
\end{deluxetable*}

\end{document}